\documentclass[twocolumn]{aastex701}
\usepackage{amsmath}
\usepackage{color}
\usepackage{booktabs}
\def\beq{\begin{eqnarray}}
\def\eeq{\end{eqnarray}}

\begin{document}

\title{Statistical analysis of multi-band plateaus in gamma-ray burst afterglows}

\author[0000-0002-6588-2652]{Xiao-Yan Li}
\affiliation{Department of Astronomy, Xiamen University, Xiamen, Fujian 361005, China; tongliu@xmu.edu.cn}
\email{lixiaoyan@stu.xmu.edu.cn}

\author[0000-0001-8678-6291]{Tong Liu}
\affiliation{Department of Astronomy, Xiamen University, Xiamen, Fujian 361005, China; tongliu@xmu.edu.cn}
\email{tongliu@xmu.edu.cn}

\author[0000-0002-4448-0849]{Bao-Quan Huang}
\affiliation{College of Intelligent Manufacturing, Nanning University, Nanning, Guangxi 530299, China; huangbaoquan@unn.edu.cn}
\email{huangbaoquan@unn.edu.cn}

\author[0000-0002-2191-7286]{Chen Deng}
\affiliation{School of Astronomy and Space Science, Nanjing University, Nanjing, Jiangsu 210023, China}
\email{dengchen@nju.edu.cn}

\begin{abstract}
Plateau features are frequently observed in the afterglows of gamma-ray bursts (GRBs), yet their physical origins remain under debate. In this work, we compile a sample of 124 GRBs with known redshifts and simultaneous X-ray and optical afterglow observations. We categorize them into four subsets based on the existence of plateaus and the bands in which they appear. Namely, Dataset 1: plateaus are detected simultaneously in both X-ray and optical bands (75 bursts); Dataset 2: plateaus are only in X-rays (15 bursts); Dataset 3: plateaus appear only in the optical (17 bursts); Dataset 4: no plateaus in either band (17 bursts). We employ these datasets to test the applicability of the energy-injection model by examining whether the temporal decay index $\alpha$ and the spectral index $\beta$ of GRB afterglows simultaneously satisfy the closure relations in X-ray and optical bands. We find that 47 bursts of Dataset 1 simultaneously obey the closure relations in both bands under the conditions of the electron spectral index $p>2$ and the injection parameter $q\in (0, 0.5)$, and 69 of the dataset for $p>1$ and $q\in (0, 0.8)$, providing a strong support for the energy-injection interpretation. However, for Datasets 2 and 3, although $\alpha$ and $\beta$ of the plateaus mostly satisfy the closure relations, those in the other band show significant deviations, which implies that bursts with a single-band plateau are inconsistent with the interpretation of energy injection. Furthermore, we also compare the isotropic X-ray energy of plateaus with the rotational energy budget of millisecond magnetars.
\end{abstract}


\section{Introduction}

Gamma-ray bursts (GRBs) are among the most energetic transients known in the Universe. Observationally, GRBs generally appear as brief and intense $\gamma$-rays, i.e., the prompt emission, followed by a long-lasting afterglow emission across multiple wavelengths from radio to X-ray. Based on the prompt duration $T_{90}$, GRBs are classified as long-duration GRBs (LGRBs; $T_{90}>2\,\mathrm{s}$) and short-duration GRBs (SGRBs; $T_{90}<2\, \mathrm{s}$) \citep[e.g.,][]{1993ApJ...413L.101K}. The progenitor of LGRBs is generally believed to be collapsars, as indicated by their association with Type Ib/c supernovae \citep[e.g.,][]{1993ApJ...405..273W,1998ApJ...494L..45P,2003Natur.423..847H,2003ApJ...591L..17S,2004ApJ...609L...5M,2011ApJ...743..204B}, although supernova-less LGRBs have also been reported, while mergers of binary neutron stars (NSs) or NS–black hole (BH) are thought to be the progenitor of SGRBs \citep[e.g.,][]{1989Natur.340..126E,1991AcA....41..257P,1992ApJ...395L..83N,2007PhR...442..166N,2017ApJ...848L..13A,2019FrPhy..1464402Z}. 
The central engines of GRBs could be either a stellar-mass BH surrounded by a hyperaccretion disk \citep[e.g.,][]{2017NewAR..79....1L}
or a massive millisecond magnetar \citep[e.g.,][]{1992ApJ...392L...9D,1992Natur.357..472U,1998PhRvL..81.4301D,1998AA...333L..87D,2011MNRAS.413.2031M}. 
  
With the observation of \textit{Swift} satellite, the diversity of light curves of GRB afterglows has been revealed \citep[e.g.,][]{2004ApJ...611.1005G,2005SSRv..120..165B}. A canonical light curve consisting of five components for X-ray afterglows thus was proposed: the initial steep decay, the shallow decay (hereafter referred to as plateaus) with a typical slope $\sim -0.5$, the normal decay which is consistent with the standard afterglow component, the late steep decay which is often called post-jet-break phase, and X-ray flares \citep[e.g.,][]{2006ApJ...642..389N,2006ApJ...647.1213O,2006ApJ...642..354Z,2007ApJ...666.1002Z}. In addition, \cite{2012ApJ...758...27L} also identified eight possible emission components for the light curves of optical afterglows. These include: the prompt optical flares tracking the $\gamma$-ray emission, the early optical flashes which should be relative to the external reverse shocks, the plateau phase, the normal decay, the post-jet-break phase, optical flares, re-brightening features observed in some GRBs, and supernova bumps. For radio afterglows, the typically low fluxes make them challenging to observe, resulting in a detection ratio of approximately $6.6\%$ \citep[e.g.,][]{2012ApJ...746..156C}.

Among the diverse features observed in GRB afterglow light curves, the plateaus are relatively common and typically appear concurrently across multiple bands. A promising interpretation is the energy-injection model in which the spin-down luminosity of a magnetar \citep[e.g.,][]{2001ApJ...552L..35Z, 2013ApJ...779L..25F} or a supramassive, fast-rotating quark star \citep[e.g.,][]{2016PhRvD..94h3010L,2018ApJ...854..104H} continuously injects energy into the initial fireball to refresh the forward shock \citep[e.g.,][]{2006ApJ...642..354Z,2006ApJ...642..389N,2006MNRAS.369.2059P,2006MNRAS.366.1357P}, and the restart of long-lived BH hyperaccretion systems cannot be ruled out \citep[e.g.,][]{2017NewAR..79....1L}. In addition, various models have also been proposed, including off-axis observations \citep[e.g.,][]{2006ApJ...641L...5E,2007ApJ...655..973K}, evolving microphysics \citep[e.g.,][]{2006AA...458....7I,2006MNRAS.369..197F}, structured jet effects \citep[e.g.,][]{2006MNRAS.370.1946G,2007ApJ...656L..57J,2021ApJ...922...22L}, a two-component ejecta \citep[e.g.,][]{2006ApJ...640L.139T,2009ApJ...690L.118Y}, a delayed deceleration jet \citep{2015ApJ...806..205D}, and a precessing jet \citep[e.g.,][]{2021ApJ...916...71H}. However, the plateaus sometimes exhibit chromatic behavior, namely that they appear only in a single band or terminate at different times in different bands \citep[e.g.,][]{2006MNRAS.366.1357P,2006MNRAS.369..197F,2007ApJ...670..565L,2011MNRAS.414.3537P}. This may challenge the energy injection scenario.

Therefore, in this work, combining with a sample of GRBs with simultaneous X-ray and optical afterglow observations, we seek to test the applicability of the energy-injection model by the closure relations derived in the framework of the energy injection. This paper is organized as follows. In Section 2, the collection and analysis of the sample are presented. The main results are shown in Section 3. Conclusions and discussion are presented in Section 4.

\section{Sample Selection and Data Analysis} 

\subsection{Sample Selection}

We collect GRBs with afterglows observed simultaneously in both X-ray and optical bands, identified redshift $z$, and measured $T_{90}$. Most of them are obtained from the literature \citep[e.g.,][]{2015ApJS..219....9W,2015ApJ...805...13L}.  Moreover, to maximize the size of the sample, we further searched the \textit{Swift} archive and the GCN Circulars. The adopted criterion is that there must be more than 5 data points in the optical and X-ray bands, respectively. Finally, we obtain a sample of 126 GRBs spanning the time range of March 2005 to October 2022. It is worth noting that the radio observations are relatively sparse, with only 33/125 cases detected in our sample. Moreover, among them, only GRB~071003 exhibits a radio plateau, but there are no plateaus in other bands. Therefore, the radio behavior is not considered in this work.

Generally, the plateaus refer to a distinct phase in the afterglow light curve during which the flux remains nearly constant and decays more slowly than the normal decay phase. In this work, we adopt a plateau as a light curve segment with the temporal power-law decay index in the range $-1 \leq \alpha \leq 1$ \citep[e.g.,][]{2019ApJS..245....1T,2023ApJ...943..126D}. For the bursts collected from the literature, we directly adopt the reported temporal decay indices. For the remaining bursts, we perform a fit by using either a single power-law (SPL):
\begin{equation}
F(t) = F_0 \, t^{-\alpha},
\end{equation}
or a smoothly broken power-law (BPL) \cite[e.g.,][]{2015ApJS..219....9W,2015ApJ...805...13L}:
\begin{equation}
F(t) = F_0 \left[ \left( \frac{t}{T_{\rm b}} \right)^{\alpha_1 \omega} + \left( \frac{t}{T_{\rm b}} \right)^{\alpha_2 \omega} \right]^{-1/\omega},
\label{eq:bkpl}
\end{equation}
where $\alpha$ is the temporal decay slope for the SPL model and $\alpha_1$ ($\alpha_2$) correspond to the temporal slopes before (after) the break time $T_{\rm b}$ for the BPL model. The flux at the break time is given by $F_{\rm b} = F_0 \, 2^{-1/\omega}$, and $\omega$ quantifies the sharpness of the break. In our analysis, we fix $\omega = 1$ for all BPL fits. To ensure consistency across the sample, we refitted the bursts whose temporal indices had been derived only from a single band in previous literature. To identify the best description of the light curves, we performed a comparative analysis using the Joint BPL, PL+PL, and PL+BPL models, and we adopted the model with the minimized  Akaike Information Criterion (AIC) \cite[e.g.,][]{1974ITAC...19..716A}.

Interestingly, we notice that in some cases the plateaus are chromatic, which poses a challenge for the standard energy-injection interpretation. To examine the general applicability of the energy-injection model, we classify our GRB sample into four categories:  

(1) Dataset 1: plateaus detected simultaneously in the light curves of optical and X-ray afterglows (75 bursts, see Table 1); 

(2) Dataset 2: the plateaus detected only in the light curves of X-ray afterglows (15 bursts, see Table 2);  

(3) Dataset 3: plateaus detected only in the light curves of optical afterglows (17 bursts, see Table 3); 

(4) Dataset 4: no plateaus detected in either the light curves of optical or X-ray afterglows (17 bursts, see Table 4).    

\begingroup
\def\tablecomments#1{\par\smallskip\noindent\footnotesize #1\par}
\newcommand{\ssgap}{0ex}
\newcommand{\headroom}{\rule{0pt}{4.5ex}} 
\newcommand{\headunit}[2]{\shortstack[c]{\headroom #1\\[\ssgap]\scriptsize(#2)}}
\newcommand{\headblank}[1]{\shortstack[c]{\headroom #1\\[\ssgap]\scriptsize\vphantom{(ks)}}}

\startlongtable
\begin{deluxetable}{lccccccccccccc} 
\tabletypesize{\scriptsize}
\tablecaption{Dataset 1: Plateaus detected in both X-ray and optical bands\label{MyTabA}}
\tablewidth{0pt}
\tablehead{
  \colhead{\headblank{GRB}} &
  \colhead{\headblank{$z$}} &
  \colhead{\headunit{$T_{\rm 90}$}{s}} &
  \colhead{\headunit{$T_{\rm b,X}$}{ks}} &
  \colhead{\headblank{$\alpha_{\rm X,1}$}} &
  \colhead{\headblank{$\alpha_{\rm X,2}$}} &
  \colhead{\headblank{$\beta_{\rm X}$}} &
  \colhead{\headunit{$T_{\rm b,o}$}{ks}} &
  \colhead{\headblank{$\alpha_{\rm o,1}$}} &
  \colhead{\headblank{$\alpha_{\rm o,2}$}} &
  \colhead{\headblank{$\beta_{\rm o}$}} &
  \colhead{\headblank{Radio?}} &
  \colhead{\headblank{Environment}} &
  \colhead{\headblank{Ref}}
}
\startdata
050319 & 3.24 & 15 & 55 & $0.58^{+\,0.07}_{-\,0.07}$ & $1.72^{+\,0.11}_{-\,0.11}$ & $1.01^{+\,0.07}_{-\,0.07}$ & - & $0.39^{+\,0.06}_{-\,0.06}$ & $1.02^{+\,0.04}_{-\,0.04}$ & $0.74^{+\,0.42}_{-\,0.42}$ & ~ & ISM/Wind  & 1 \\ 
050401 & 2.90 & 38 & 4.3 & $0.76^{+\,0.05}_{-\,0.05}$ & $1.67^{+\,0.11}_{-\,0.11}$ & $0.79^{+\,0.13}_{-\,0.13}$ & - & $0.5^{+\,0.08}_{-\,0.08}$ & $0.89^{+\,0.08}_{-\,0.08}$ & $0.5^{+\,0.2}_{-\,0.2}$ & y & ISM*/Wind  & 1 \\ 
050408 & 1.24 & 34 & 40.7 & $0.73^{+\,0.15}_{-\,0.15}$ & $1.17^{+\,0.22}_{-\,0.22}$ & $1.14^{+\,0.14}_{-\,0.14}$ & - & $0.49^{+\,0.01}_{-\,0.01}$ & $1.29^{+\,0.11}_{-\,0.11}$ & $0.28^{+\,0.33}_{-\,0.33}$ & ~ & ISM*  & 1 \\ 
050416A & 0.65 & 2.4 & 11.0 & $0.66^{+\,0.12}_{-\,0.12}$ & $0.99^{+\,0.02}_{-\,0.02}$ & $1.07^{+\,0.11}_{-\,0.11}$ & - & $0.26^{+\,0.07}_{-\,0.07}$ & $1.12^{+\,0.12}_{-\,0.12}$ & 1.30 & y & ISM/Wind  & 1 \\ 
050730 & 3.97 & 155 & 90.1 & $0.45^{+\,0.13}_{-\,0.13}$ & $2.64^{+\,0.20}_{-\,0.20}$ & $1.62^{+\,0.04}_{-\,0.04}$ & - & $0.48^{+\,0.05}_{-\,0.05}$ & $1.47^{+\,0.06}_{-\,0.06}$ & $0.52^{+\,0.05}_{-\,0.05}$ & y &   & 1 \\ 
050801 & 1.38 & 20 & 0.2 & $0.24^{+\,0.11}_{-\,0.11}$ & $1.18^{+\,0.03}_{-\,0.03}$ & $0.92^{+\,0.17}_{-\,0.17}$ & - & $0.07^{+\,0.01}_{-\,0.01}$ & $1.20^{+\,0.01}_{-\,0.01}$ & $0.69^{+\,0.34}_{-\,0.34}$ & ~ & ISM/Wind  & 1 \\ 
050824 & 0.83 & 25 & 37.5 & $0.04^{+\,0.32}_{-\,0.32}$ & $0.98^{+\,0.17}_{-\,0.17}$ & $0.96^{+\,0.07}_{-\,0.07}$ & \dots & $0.66^{+\,0.01}_{-\,0.01}$ & \dots & $0.45^{+\,0.18}_{-\,0.18}$ & ~ & ISM*/Wind & 2 \\ 
050922C & 2.20 & 5 & 3.2 & $0.97^{+\,0.04}_{-\,0.04}$ & $1.49^{+\,0.05}_{-\,0.05}$ & $1.17^{+\,0.11}_{-\,0.11}$ & $6.4$ & $0.46^{+\,0.03}_{-\,0.03}$ & $1.50^{+\,0.02}_{-\,0.02}$ & $0.51^{+\,0.05}_{-\,0.05}$ & ~ &  ISM* & 3 \\ 
051109A & 2.35 & 25 & 3.5 & $0.24^{+\,0.04}_{-\,0.04}$ & $1.22^{+\,0.11}_{-\,0.11}$ & $0.98^{+\,0.08}_{-\,0.08}$ & - & $0.64^{+\,0.08}_{-\,0.08}$ & $1.07^{+\,0.12}_{-\,0.12}$ & $0.70^{+\,0.05}_{-\,0.05}$ & y & ISM*/Wind  & 1 \\  
051221A & 0.65 & 2.4 & 25.1 & $0.35^{+\,0.08}_{-\,0.08}$ & $1.34^{+\,0.04}_{-\,0.04}$ & $1.06^{+\,0.14}_{-\,0.14}$ & - & $0.34^{+\,0.07}_{-\,0.07}$ & $1.24^{+\,0.04}_{-\,0.04}$ & $0.64^{+\,0.05}_{-\,0.05}$ & y & ISM  & 1 \\ 
060206 & 4.06 & 11 & 12.5 & $0.40^{+\,0.09}_{-\,0.09}$ & $1.50^{+\,0.06}_{-\,0.06}$ & $1.20^{+\,0.31}_{-\,0.31}$ & - & $0.42^{+\,0.09}_{-\,0.09}$ & $1.43^{+\,0.10}_{-\,0.10}$ & $0.73^{+\,0.05}_{-\,0.05}$ & n & ISM & 1 \\ 
060210 & 3.91 & 255 & 5.0 & $0.53^{+\,0.06}_{-\,0.06}$ & $1.30^{+\,0.12}_{-\,0.12}$ & $1.08^{+\,0.08}_{-\,0.08}$ & - & $0.53^{+\,0.05}_{-\,0.05}$ & $1.77^{+\,0.14}_{-\,0.14}$ & $0.37^{+\,0.08}_{-\,0.08}$ & n &  & 1 \\ 
060526 & 3.22 & 13.8 & 50.1 & $0.67^{+\,0.08}_{-\,0.08}$ & $2.06^{+\,0.15}_{-\,0.15}$ & $0.90^{+\,0.11}_{-\,0.11}$ & - & $0.56^{+\,0.10}_{-\,0.10}$ & $1.93^{+\,0.10}_{-\,0.10}$ & $0.51^{+\,0.32}_{-\,0.32}$ & ~ & ISM*/Wind*  & 1 \\ 
060605 & 3.77 & 15 & 15.0 & $0.55^{+\,0.08}_{-\,0.08}$ & $2.82^{+\,0.15}_{-\,0.15}$ & $1.02^{+\,0.09}_{-\,0.09}$ & - & $0.13^{+\,0.09}_{-\,0.09}$ & $2.64^{+\,0.15}_{-\,0.15}$ & 1.06 & n & ISM/Wind  & 1 \\ 
060607A & 3.07 & 100 & 9.5 & $0.36^{+\,0.03}_{-\,0.03}$ & $3.10^{+\,0.12}_{-\,0.12}$ & $0.62^{+\,0.06}_{-\,0.06}$ & - & -0.93 & 4.6 & $0.72^{+\,0.27}_{-\,0.27}$ & ~ &   & 1 \\ 
060614 & 0.13 & 102 & 44.0 & $0.11^{+\,0.10}_{-\,0.10}$ & $1.97^{+\,0.15}_{-\,0.15}$ & $0.90^{+\,0.09}_{-\,0.09}$ & - & $-0.35^{+\,0.04}_{-\,0.04}$ & $1.90^{+\,0.12}_{-\,0.12}$ & $0.47^{+\,0.04}_{-\,0.04}$ & ~ & ISM  & 1 \\ 
$060708^{\rm s}$ & 9.8 & 1.92 & 8.9 & $0.60^{+\,0.05}_{-\,0.05}$ & $1.32^{+\,0.05}_{-\,0.05}$ & $1.16^{+\,0.2}_{-\,0.2}$ & $0.7$ & $0.06^{+\,0.10}_{-\,0.10}$ & $0.85^{+\,0.03}_{-\,0.03}$ & $0.88^{+\,0.05}_{-\,0.05}$ & ~ & ISM  & 3 \\ 
060714 & 2.71 & 115 & 5.9 & $0.48^{+\,0.09}_{-\,0.09}$ & $1.34^{+\,0.11}_{-\,0.11}$ & $1.10^{+\,0.19}_{-\,0.19}$ & - & $0.15^{+\,0.06}_{-\,0.06}$ & $1.04^{+\,0.15}_{-\,0.15}$ & $0.44^{+\,0.04}_{-\,0.04}$ & ~ &  & 1 \\ 
060729 & 0.54 & 116 & 53.0 & $0.05^{+\,0.01}_{-\,0.01}$ & $1.45^{+\,0.11}_{-\,0.11}$ & $1.02^{+\,0.04}_{-\,0.04}$ & - & $0.10^{+\,0.05}_{-\,0.05}$ & $1.40^{+\,0.15}_{-\,0.15}$ & $0.78^{+\,0.03}_{-\,0.03}$ & ~ & ISM  & 1 \\ 
060906 & 3.69 & 43.6 & 12.6 & $0.29^{+\,0.06}_{-\,0.06}$ & $1.82^{+\,0.17}_{-\,0.17}$ & $1.07^{+\,0.21}_{-\,0.21}$ & \dots & $0.83^{+\,0.17}_{-\,0.17}$ & \dots & $0.56^{+\,0.02}_{-\,0.02}$ & ~ & Wind & 3 \\ 
060908 & 1.88 & 19.3 & 1.1 & $0.54^{+\,0.05}_{-\,0.05}$ & $1.66^{+\,0.05}_{-\,0.05}$ & $1.13^{+\,0.19}_{-\,0.19}$ & - & $0.71^{+\,0.04}_{-\,0.04}$ & $1.11^{+\,0.09}_{-\,0.09}$ & $0.24^{+\,0.2}_{-\,0.2}$ & n &  & 1 \\ 
061021 & 0.35 & 46 & 16.6 & $0.60^{+\,0.03}_{-\,0.03}$ & $1.16^{+\,0.02}_{-\,0.02}$ & $0.98^{+\,0.1}_{-\,0.1}$ & $88.8$ & $0.65^{+\,0.01}_{-\,0.01}$ & $2.01^{+\,0.24}_{-\,0.24}$ & $0.55^{+\,0.02}_{-\,0.02}$ & ~ & Wind & 3 \\ 
061121 & 1.31 & 81 & 7.7 & $0.34^{+\,0.03}_{-\,0.03}$ & $1.55^{+\,0.03}_{-\,0.03}$ & $0.93^{+\,0.02}_{-\,0.02}$ & $40.9$ & $0.53^{+\,0.08}_{-\,0.08}$ & $1.26^{+\,0.14}_{-\,0.14}$ & 0.6 & y & ISM  & 2 \\ 
070110 & 2.35 & 85 & 20.3 & $0.30^{+\,0.10}_{-\,0.10}$ & $5.10^{+\,0.30}_{-\,0.30}$ & $1.08^{+\,0.11}_{-\,0.11}$ & - & $0.16^{+\,0.06}_{-\,0.06}$ & $1.67^{+\,0.12}_{-\,0.12}$ & $0.55^{+\,0.04}_{-\,0.04}$ & ~ & ISM & 1 \\ 
070208 & 1.17 & 48 & 9.0 & $0.43^{+\,0.03}_{-\,0.03}$ & $1.78^{+\,0.13}_{-\,0.13}$ & $1.20^{+\,0.20}_{-\,0.20}$ & \dots & $0.49^{+\,0.06}_{-\,0.06}$ & \dots & 0.68 & ~ & ISM/Wind  & 1 \\ 
070518 & 1.16 & 5.5 & 40.1 & $0.41^{+\,0.06}_{-\,0.06}$ & $1.51^{+\,0.09}_{-\,0.09}$ & $1.20^{+\,0.34}_{-\,0.34}$ & - & $0.70^{+\,0.07}_{-\,0.07}$ & $1.80^{+\,0.11}_{-\,0.11}$ & 0.80 & n &  ISM* & 1 \\ 
070802 & 2.45 & 16.4 & 9.5 & 0.21 & 1.36 & 1.4 & \dots & 0.61 & \dots & $1.07^{+\,0.31}_{-\,0.31}$ & ~ &   & 4 \\ 
071010B & 0.95 & 35.7 & \dots & 0.69 & \dots & $0.92^{+0.32}_{-0.30}$ & \dots & 0.58 & \dots & $0.68^{+\,0.14}_{-\,0.14}$ & y & ISM*/Wind  & 4 \\ 
071031 & 2.69 & 180 & \dots & $0.82^{+\,0.05}_{-\,0.05}$ & \dots & $0.71^{+\,0.14}_{-\,0.14}$ & \dots & $0.79^{+\,0.05}_{-\,0.05}$ & \dots & $0.64^{+\,0.05}_{-\,0.05}$ & ~ & Wind  & 1 \\ 
071112C & 0.82 & 15 & 1.5 & $0.50^{+\,0.08}_{-\,0.08}$ & $1.49^{+\,0.11}_{-\,0.11}$ & $0.67^{+\,0.13}_{-\,0.13}$ & - & $0.30^{+\,0.07}_{-\,0.07}$ & $0.95^{+\,0.11}_{-\,0.11}$ & $0.63^{+\,0.29}_{-\,0.29}$ & n & ISM  & 1 \\ 
080310 & 2.43 & 365 & 5.1 & $0.03^{+\,0.06}_{-\,0.06}$ & $1.24^{+\,0.08}_{-\,0.08}$ & $0.95^{+\,0.18}_{-\,0.18}$ & - & $0.11^{+\,0.01}_{-\,0.01}$ & $1.24^{+\,0.02}_{-\,0.02}$ & $0.42^{+\,0.12}_{-\,0.12}$ & ~ & ISM  & 1 \\ 
080319B & 0.94 & 50 & 3.0 & $0.73^{+\,0.05}_{-\,0.05}$ & $2.73^{+\,0.16}_{-\,0.16}$ & $0.81^{+\,0.07}_{-\,0.07}$ & - & $0.93^{+\,0.11}_{-\,0.11}$ & $1.60^{+\,0.12}_{-\,0.12}$ & $0.51^{+\,0.26}_{-\,0.26}$ & y & Wind  & 1 \\ 
080413B & 1.10 & 8 & 148.5 & $0.92^{+\,0.16}_{-\,0.16}$ & $1.91^{+\,0.23}_{-\,0.23}$ & $0.94^{+\,0.07}_{-\,0.07}$ & - & $0.31^{+\,0.15}_{-\,0.15}$ & $1.89^{+\,0.22}_{-\,0.22}$ & $0.25^{+\,0.07}_{-\,0.07}$ & n &  & 1 \\ 
080603A & 1.69 & 180 & \dots & $0.96^{+\,0.05}_{-\,0.05}$ & \dots & $1.01^{+\,0.10}_{-\,0.10}$ & \dots & $0.95^{+\,0.03}_{-\,0.03}$ & \dots & $0.98^{+\,0.04}_{-\,0.04}$ & y & ISM*/Wind  & 1 \\ 
080605 & 1.64 & 18 & 0.6 & $0.40^{+\,0.08}_{-\,0.08}$ & $1.50^{+\,0.03}_{-\,0.03}$ & $0.71^{+\,0.02}_{-\,0.02}$ & \dots & $0.64^{+\,0.01}_{-\,0.01}$ & \dots & $0.58^{+\,0.35}_{-\,0.35}$ & ~ & ISM & 2 \\ 
080710 & 0.85 & 120 & 6.8 & $0.34^{+\,0.04}_{-\,0.04}$ & $1.57^{+\,0.14}_{-\,0.14}$ & $1.00^{+\,0.11}_{-\,0.11}$ & - & $0.39^{+\,0.05}_{-\,0.05}$ & $1.32^{+\,0.11}_{-\,0.11}$ & $0.80^{+\,0.09}_{-\,0.09}$ & ~ & ISM & 1 \\ 
081007 & 0.53 & 9.7 & 113.4 & $0.68^{+\,0.05}_{-\,0.05}$ & $1.56^{+\,0.17}_{-\,0.17}$ & $0.87^{+\,0.05}_{-\,0.05}$ & \dots & $0.73^{+\,0.01}_{-\,0.01}$ & \dots & $0.27^{+\,0.11}_{-\,0.11}$ & y &  & 2 \\ 
081008 & 1.97 & 185.5 & 9.5 & $0.87^{+\,0.15}_{-\,0.15}$ & $1.68^{+\,0.08}_{-\,0.08}$ & $0.98^{+\,0.11}_{-\,0.11}$ & - & $0.64^{+\,0.06}_{-\,0.06}$ & $1.60^{+\,0.09}_{-\,0.09}$ & $0.40^{+\,0.23}_{-\,0.23}$ & n & ISM*/Wind  & 1 \\ 
$090426^{\rm s}$ & 2.6 & 1.2 & 0.2 & $0.13^{+\,0.02}_{-\,0.02}$ & $1.04^{+\,0.05}_{-\,0.05}$ & $1.03^{+\,0.15}_{-\,0.15}$ & - & $0.14^{+\,0.09}_{-\,0.09}$ & $1.25^{+\,0.04}_{-\,0.04}$ & $0.76^{+\,0.14}_{-\,0.14}$ & ~ & ISM & 1 \\ 
090529A & 2.6 & 100 & 97.5 & $0.52^{+0.06}_{-0.06}$ & $1.09^{+\,0.20}_{-\,0.20}$ & $0.48^{+0.67}_{-0.23}$ & - & $0.45^{+\,0.03}_{-\,0.03}$ & $0.90^{+\,0.07}_{-\,0.07}$ & $0.1^{+\,0.46}_{-\,0.46}$ & ~ & ISM* & 5 \\ 
090618 & 0.54 & 113.2 & 45.1 & $0.93^{+\,0.09}_{-\,0.09}$ & $1.74^{+\,0.10}_{-\,0.10}$ & $0.92^{+\,0.05}_{-\,0.05}$ & - & $0.76^{+\,0.11}_{-\,0.11}$ & $1.53^{+\,0.11}_{-\,0.11}$ & $0.50^{+\,0.05}_{-\,0.05}$ & y & Wind  & 1 \\ 
091018 & 0.97 & 4.4 & 0.3 & $-0.17^{+\,0.26}_{-\,0.26}$ & $1.25^{+\,0.03}_{-\,0.03}$ & $0.92^{+\,0.03}_{-\,0.03}$ & 125.1 & $0.91^{+\,0.01}_{-\,0.01}$ & $2.84^{+\,0.32}_{-\,0.32}$ & $0.61^{+\,0.02}_{-\,0.02}$ & ~ &   & 2 \\ 
091029 & 2.75 & 39.2 & 20.8 & $0.32^{+\,0.09}_{-\,0.09}$ & $1.35^{+\,0.08}_{-\,0.08}$ & $1.12^{+\,0.08}_{-\,0.08}$ & - & $0.48^{+\,0.06}_{-\,0.06}$ & $1.34^{+\,0.09}_{-\,0.09}$ & $0.49^{+\,0.12}_{-\,0.12}$ & ~ & ISM*/Wind & 1 \\ 
091127 & 0.49 & 7.1 & 35.3 & $0.96^{+\,0.05}_{-\,0.05}$ & $1.59^{+\,0.12}_{-\,0.12}$ & $0.68^{+\,0.11}_{-\,0.11}$ & - & $0.55^{+\,0.11}_{-\,0.11}$ & $1.50^{+\,0.11}_{-\,0.11}$ & 0.18 & n &  & 1 \\ 
100219A & 4.67 & 18.8 & 1.8 & $0.54^{+\,0.07}_{-\,0.07}$ & $1.65^{+\,0.15}_{-\,0.15}$ & $0.69^{+\,0.23}_{-\,0.23}$ & - & $0.74^{+\,0.08}_{-\,0.08}$ & $1.91^{+\,0.12}_{-\,0.12}$ & 0.56 & ~ & Wind  & 1 \\ 
100418A & 0.62 & 7 & 90.1 & $-0.12^{+\,0.03}_{-\,0.03}$ & $1.57^{+\,0.11}_{-\,0.11}$ & $1.04^{+\,0.29}_{-\,0.29}$ & - & $0.11^{+\,0.01}_{-\,0.01}$ & $1.60^{+\,0.10}_{-\,0.10}$ & $0.98^{+\,0.09}_{-\,0.09}$ & y & ISM  & 1 \\ 
100621A & 0.54 & 63.6 & 80.5 & $0.68^{+\,0.03}_{-\,0.03}$ & $1.69^{+\,0.09}_{-\,0.09}$ & $1.37^{+\,0.03}_{-\,0.03}$ & 40 & $0.22^{+\,0.02}_{-\,0.02}$ & $2.41^{+\,0.18}_{-\,0.18}$ & $0.78^{+\,0.09}_{-\,0.09}$ & ~ & ISM  & 2 \\ 
100814A & 1.44 & 177 & 215.3 & $0.50^{+\,0.02}_{-\,0.02}$ & $2.38^{+\,0.09}_{-\,0.09}$ & $0.84^{+\,0.02}_{-\,0.02}$ & 391.4 & $0.09^{+\,0.01}_{-\,0.01}$ & $3.90^{+\,0.15}_{-\,0.15}$ & $0.41^{+\,0.04}_{-\,0.04}$ & y &   & 2 \\ 
110213A & 1.46 & 48 & 3.1 & $-0.19^{+\,0.06}_{-\,0.06}$ & $1.90^{+\,0.03}_{-\,0.03}$ & $1.04^{+\,0.02}_{-\,0.02}$ & 15.2 & $0.20^{+\,0.01}_{-\,0.01}$ & $2.01^{+\,0.02}_{-\,0.02}$ & $0.9^{+\,0.07}_{-\,0.07}$ & ~ &   & 2 \\ 
110715A & 0.82 & 13 & 0.1 & $-0.80^{+\,0.32}_{-\,0.32}$ & $1.00^{+\,0.01}_{-\,0.01}$ & $0.85^{+\,0.02}_{-\,0.02}$ & 374.9 & $0.52^{+\,0.01}_{-\,0.01}$ & $2.87^{+\,0.30}_{-\,0.30}$ & $0.63^{+\,0.28}_{-\,0.28}$ & y &   & 2 \\ 
111228A & 0.71 & 101 & 7.5 & $0.19^{+\,0.07}_{-\,0.07}$ & $1.28^{+\,0.04}_{-\,0.04}$ & $0.97^{+\,0.03}_{-\,0.03}$ & 19.1 & $0.25^{+\,0.03}_{-\,0.03}$ & $1.54^{+\,0.04}_{-\,0.04}$ & $0.69^{+\,0.07}_{-\,0.07}$ & ~ & ISM  & 2 \\ 
120815A & 2.36 & 9.7 & \dots & $0.86^{+\,0.06}_{-\,0.06}$ & \dots & $0.72^{+\,0.11}_{-\,0.11}$ & \dots & $0.63^{+\,0.04}_{-\,0.04}$ & \dots & $0.78^{+\,0.01}_{-\,0.01}$ & ~ & ISM*  & 1 \\ 
121217A & 3.1 & 778 & 24.9 & $0.47^{+\,0.09}_{-\,0.09}$ & $1.43^{+\,0.05}_{-\,0.05}$ & $1.06^{+\,0.16}_{-\,0.16}$ & \dots & $0.78^{+\,0.01}_{-\,0.01}$ & \dots & $0.87^{+\,0.04}_{-\,0.04}$ & ~ & ISM* & 3 \\ 
140423A & 3.26 & 134.0 & 26.7 & $1.00^{+0.04}_{-0.04}$ & $1.45^{+0.08}_{-0.08}$ & $0.94^{+0.14}_{-0.13}$ & \dots & $0.99^{+0.02}_{-0.02}$ & \dots & $0.95^{+0.08}_{-0.08}$ & & ISM*/Wind & 5  \\ 
140430A & 1.6 & 173.6 & 39.25 & $0.42^{+0.08}_{-0.07}$ & $1.44^{+0.22}_{-0.23}$ & $1.13^{+0.15}_{-0.15}$ & - & $0.48^{+0.02}_{-0.02}$ & $1.64^{+0.10}_{-0.12}$ & $0.65^{+\,0.06}_{-\,0.06}$ & ~ & ISM*  & 5 \\ 
140506A & 0.89 & 111.1 & 10.3 & $0.78^{+0.05}_{-0.05}$ &  $0.96^{+0.02}_{-0.02}$ & $0.75^{+0.08}_{-0.08}$ & \dots & $0.84^{+0.02}_{-0.02}$ & \dots & $0.24^{+\,0.27}_{-\,0.27}$ & ~ & Wind  & 5 \\ 
140518A & 4.7 & 60.5 & 4.0 & $0.45^{+0.05}_{-0.05}$ & $1.93^{+0.13}_{-0.13}$ & $0.91^{+0.12}_{-0.12}$ & - & $-0.02^{+0.05}_{-0.05}$ & $1.06^{+0.07}_{-0.07}$ & $0.22^{+0.10}_{-0.34}$ & ~ &   & 5 \\ 
140703A & 3.14 & 67.1 & 20.4 & $0.91^{+0.06}_{-0.06}$ & $2.35^{+0.12}_{-0.12}$ & $0.81^{+0.09}_{-0.09}$ & - & $-0.1^{+0.07}_{-0.07}$ & $1.65^{+0.10}_{-0.10}$ & $0.81^{+0.07}_{-0.07}$ & y & ISM*  & 5 \\ 
$140903A^{\rm s}$ & 0.35 & 0.3 & 8.1 & $0.17^{+0.04}_{-0.04}$ & $1.15^{+0.07}_{-0.07}$ & $0.59^{+0.13}_{-0.13}$ & \dots & $0.05^{+0.03}_{-0.03}$ & \dots & $0.72^{+0.05}_{-0.05}$ & y & ISM  & 5 \\ 
140907A & 1.21 & 79.2 & \dots & $0.98^{+0.03}_{-0.03}$ & \dots & $1.12^{+0.14}_{-0.14}$ & 68.7 & $0.70^{+0.06}_{-0.06}$ & $1.70^{+0.29}_{-0.29}$ & $1.20^{+0.12}_{-0.12}$ & ~ & ISM/Wind  & 5 \\ 
150323A & 0.59 & 149.6 & 10.5 & $0.46^{+0.06}_{-0.06}$ & $1.17^{+0.08}_{-0.08}$ & $1.14^{+0.28}_{-0.26}$ & \dots & $0.44^{+0.02}_{-0.02}$ & \dots & $1.10^{+0.21}_{-0.21}$ & ~ & ISM/Wind  & 5 \\ 
150424A & 3 & 91 & \dots & $0.89^{+0.03}_{-0.03}$ & \dots & $0.97^{+0.22}_{-0.20}$ & 3.4 & $-0.25^{+0.16}_{-0.16}$ & $0.23^{+0.04}_{-0.04}$ & $0.41^{+0.09}_{-0.09}$ & y & ISM & 5 \\ 
151027A & 0.81 & 129.69 & 9.81 & $0.37^{+0.02}_{-0.03}$ & $1.99^{+0.00}_{-0.02}$ & $0.98^{+0.06}_{-0.06}$ & - & $-0.3^{+0.06}_{-0.07}$ & $1.76^{+0.01}_{-0.02}$ & $0.80^{+0.10}_{-0.10}$ & y & ISM  & 6 \\ 
160227A & 2.38 & 316.5 & 95.5 & $0.58^{+0.02}_{-0.02}$ & $1.48^{+0.11}_{-0.11}$ & $0.70^{+0.09}_{-0.08}$ & \dots & $0.45^{+0.07}_{-0.07}$ & \dots & $0.70^{+0.07}_{-0.07}$ & ~ & ISM* & 5 \\ 
160314A & 0.73 & 8.73 & \dots & $0.57^{+0.06}_{-0.06}$ & \dots & $0.3^{+0.5}_{-0.4}$ & \dots & $0.31^{+0.08}_{-0.08}$ & \dots & $0.90^{+0.15}_{-0.15}$ & ~ & ISM*/Wind  & 5 \\ 
160804A & 0.74 & 144.2 & 18 & $0.50^{+0.07}_{-0.07}$ & $0.90^{+0.05}_{-0.05}$ & $0.83^{+0.12}_{-0.12}$ & \dots & $0.31^{+0.03}_{-0.03}$ & \dots & $0.90^{+0.12}_{-0.12}$ & ~ & ISM  & 5 \\ 
170113A & 1.97 & 20.66 & 3.7 & $0.43^{+0.07}_{-0.07}$ & $1.22^{+0.02}_{-0.02}$ & $0.78^{+0.07}_{-0.07}$ & \dots & $0.62^{+0.05}_{-0.05}$ & \dots & $0.80^{+0.07}_{-0.07}$ & n & ISM  & 5 \\ 
170202A & 3.65 & 46.2 & 2.25 & $-0.05^{+0.09}_{-0.09}$ & $1.14^{+0.05}_{-0.05}$ & $0.98^{+0.13}_{-0.12}$ & 4.87 & $0.92^{+0.02}_{-0.02}$ & $0.75^{+0.05}_{-0.05}$ & $0.47^{+0.12}_{-0.65}$ & ~ &  Wind & 5 \\ 
170607A & 0.56 & 23 & 15.4 & $0.35^{+0.04}_{-0.04}$ & $0.98^{+0.02}_{-0.02}$ & $0.94^{+0.07}_{-0.07}$ & - & $0.38^{+0.02}_{-0.02}$ & $0.79^{+0.12}_{-0.12}$ & $0.90^{+0.07}_{-0.07}$ & ~ & ISM  & 5 \\ 
180325A & 2.25 & 94.1 & 7.6 & $0.82^{+0.01}_{-0.01}$ & $2.34^{+0.05}_{-0.05}$ & $0.79^{+0.09}_{-0.09}$ & - & $0.05^{+0.01}_{-0.01}$ & $1.52^{+0.04}_{-0.04}$ & $0.45^{+0.01}_{-0.01}$ & ~ &  & 5 \\ 
180418A & 1.55 & 4.4 & \dots &$0.99^{+0.05}_{-0.05}$ & \dots & $0.82^{+0.32}_{-0.20}$ & \dots &  $1.00^{+0.03}_{-0.03}$ & \dots & $0.70^{+0.19}_{-0.19}$ & ~ &ISM*/Wind  & 5 \\ 
180620A & 1.2 & 23.16 & 4.39 & $-0.3^{+0.09}_{-0.09}$ & $1.99^{+0.00}_{-0.01}$ & $0.39^{+0.07}_{-0.06}$ & - & $-0.55^{+0.17}_{-0.19}$ & $0.82^{+0.04}_{-0.03}$ & 0.58 & ~ &  & 7 \\ 
190106A & 1.86 & 76.8 & 16 & 0.36 & 1.3 & $0.78^{+0.03}_{-0.03}$ & 77 & 0.63 & 1.29 & $0.78^{+0.03}_{-0.03}$ & ~ & ISM* & 8 \\  
210731A & 1.25 & 22.5 & 23.33 & $0.99^{+\,0.16}_{-\,0.16}$ & $-1.84^{+\,0.04}_{-\,0.04}$ & $1.00^{+\,0.11}_{-\,0.11}$ & 22.46 & $0.44^{+\,0.62}_{-\,0.62}$ & $-1.69^{+\,0.19}_{-\,0.19}$ & $-0.81^{+\,0.05}_{-\,0.05}$ & ~ &   & 9 \\ 
210905A & 6.32 & 24 & 61.1 & $0.73^{+0.05}_{-0.05}$ & $1.09^{+0.04}_{-0.04}$ & $0.86^{+0.13}_{-0.13}$ & - & $0.66^{+0.04}_{-0.04}$ & $0.94^{+0.02}_{-0.02}$ & $0.6^{+\,0.04}_{-\,0.04}$ & y & ISM*/Wind  & 5 \\
\enddata
\tablecomments{
\textit{Notes:} SGRBs are marked with a superscript $\rm s$. ``$\dots$'' indicates that no break is observed, while ``-'' means that the optical break time is the same as that in the X-rays. The column ``Environment'' denotes the medium in which the bursts satisfy the closure relations under the condition $p>2$, and ``$*$'' marks the bursts that satisfy the closure relations only within $q \in (0, 0.8)$. In addition, for each burst, we mark radio-afterglow detections with ``y'' and non-detections with ``n'', while the blanks represent no report. The same notation applies to the following tables.\\[0.25ex]
\textit{References:} (1) \citet{2015ApJS..219....9W}; (2) \citet{2023AA...675A.117R}; (3) \citet{2015ApJ...805...13L}; (4) \citet{2012grb..confE..75H}; (5) this work; (6) \citet{2017AA...598A..23N}; (7) \citet{2019ApJ...887..254B}; (8) \citet{2023ApJ...948...30Z}; (9) \citet{2023AA...671A.116D}.%
}
\end{deluxetable}
\endgroup
\twocolumngrid

\tabletypesize{\scriptsize}
\begin{table*}
\label{MyTabB}
\caption{Dataset 2: Plateaus detected only in the X-ray Band}
\centering
\setlength{\tabcolsep}{3pt}
\begin{tabular}{llllllllllllll}
\toprule
\hline
GRB & $z$ & $T_{\rm{90}}$ & $T_{\mathrm{b,X}}$ & $\alpha_{\mathrm{X},1}$ & $\alpha_{\mathrm{X},2}$ & $\beta_{\mathrm{X}}$ & $T_{\mathrm{b,o}}$ & $\alpha_{\mathrm{o},1}$  & $\alpha_{\mathrm{o},2}$  & $\beta_{\mathrm{o}}$ & Radio? & Environment & Ref \\
~&~ & $(\mathrm{s})$ & $(\rm{ks})$ & ~ & ~ & ~ &$(\rm{ks})$  & ~ & ~ & ~ & ~&~ &~ \\ 
\hline
060418 & 1.49 & 52 & 1.3 & $0.87^{+0.13}_{-0.13}$ & $1.57^{+0.05}_{-0.05}$ & $ 0.98^{+0.22}_{-0.22}$ & \dots & $1.27^{+0.02}_{-0.02}$ & \dots & $0.78^{+0.09}_{-0.09}$ & ~ & Wind  & 1 \\
060927 & 5.47 & 22.6 & 4.2 & $0.68^{+0.12}_{-0.12}$ & $1.99^{+0.46}_{-0.46}$ & $0.86^{+0.25}_{-0.25}$ & \dots & $1.17^{+0.04}_{-0.04}$ & \dots & $0.86^{+0.03}_{-0.03}$ & ~ & Wind  & 1 \\ 
070419A & 0.97 & 116 & \dots & $0.60^{+0.02}_{-0.02}$ & \dots & $1.20^{+0.30}_{-0.30}$ & \dots & $1.28^{+0.04}_{-0.04}$ & \dots & $0.48^{+0.48}_{-0.48}$ & ~ & Wind* & 2 \\ 
080721 & 2.59 & 16.2 & 3.1 & $0.81^{+0.01}_{-0.01}$ & $1.65^{+0.07}_{-0.07}$ & $0.94^{+0.06}_{-0.06}$ & ~ & $1.17^{+0.03}_{-0.03}$ & $1.31^{+0.05}_{-0.05}$ & $0.68^{+0.02}_{-0.02}$ & n &   & 2 \\ 
081029 & 3.85 & 275 & 18 & $0.18^{+0.10}_{-0.10}$ & $3.00^{+0.22}_{-0.22}$ & $0.86^{+0.05}_{-0.05}$ & 8.6 & $-1.31^{+0.06}_{-0.06}$ & $1.90^{+0.02}_{-0.02}$ & $0.33^{+0.05}_{-0.05}$ & ~ &   & 3 \\ 
100906A & 1.73 & 114.4 & 13.2 & $0.76^{+0.03}_{-0.03}$ & $2.12^{+0.80}_{-0.80}$ & $0.96^{+0.1}_{-0.1}$ & \dots & $1.07^{+0.02}_{-0.02}$ & \dots & $0.84^{+0.22}_{-0.22}$ & ~ & ISM*/Wind*  & 1 \\ 
120119A & 1.73 & 90.5 & 32.5 & $1.00^{+0.02}_{-0.02}$ & $2.03^{+0.18}_{-0.18}$ & $0.61^{+0.11}_{-0.11}$ & \dots & $1.36^{+0.03}_{-0.03}$ & \dots & $0.89^{+0.01}_{-0.01}$ & n & Wind  & 1 \\ 
120404A & 2.87 & 38.7 & 2.8 & $-0.16^{+0.38}_{-0.38}$ & $1.97^{+0.14}_{-0.14}$ & $0.78^{+0.06}_{-0.06}$ & 2.6 & $-1.65^{+0.11}_{-0.11}$ & $1.60^{+0.04}_{-0.04}$ & 1.02 & y &   & 3 \\ 
130702A & 0.15 & 59 & 89.3 & $0.10^{+0.61}_{-0.61}$ & $1.35^{+0.10}_{-0.10}$ & $0.81^{+0.03}_{-0.03}$ & \dots & $1.31^{+0.02}_{-0.02}$ & \dots & $0.71^{+0.02}_{-0.02}$ & y &   & 3 \\ 
140419A & 3.96 & 94.7 & 6.0 & $0.60^{+0.2}_{-0.2}$ & $1.55^{+0.11}_{-0.11}$ & $0.88^{+0.03}_{-0.03}$ & \dots & $1.06^{+0.01}_{-0.01}$ & \dots & $0.76^{+0.08}_{-0.08}$ & ~ && 3 \\ 
140512A & 0.73 & 154.8 & 7.7 & $0.67^{+0.01}_{-0.01}$ & $1.36^{+0.03}_{-0.03}$ & $0.82^{+0.06}_{-0.06}$ & \dots & $1.76^{+0.02}_{-0.02}$ & \dots & $0.86^{+\,0.01}_{-\,0.01}$ & ~ & ISM*  & 4 \\ 
150910A & 1.36 & 112.2 & 6.7 & $0.34^{+0.02}_{-0.02}$ & $2.49^{+0.06}_{-0.06}$ & $0.52^{+0.02}_{-0.02}$ & 1.1 & $-4.00^{+0.34}_{-0.34}$ & $1.26^{+0.01}_{-0.01}$ & $0.53^{+0.14}_{-0.14}$ & ~ &   & 3 \\ 
161014A & 2.82 & 18.3 & 2.1 & $0.41^{+0.07}_{-0.07}$ & $1.82^{+0.07}_{-0.07}$ & $0.79^{+0.11}_{-0.11}$ & \dots & $1.06^{+0.01}_{-0.01}$ & \dots & $0.80^{+0.11}_{-0.11}$ & n & ISM* & 4 \\ 
180728A & 0.12 & 8.7 & 21.2 & $0.69^{+0.17}_{-0.17}$ & $1.44^{+0.07}_{-0.07}$ & $0.76^{+0.02}_{-0.02}$ & \dots & $1.08^{+0.01}_{-0.01}$ & \dots & $0.67^{+0.05}_{-0.05}$ & ~ & Wind  & 3 \\ 
210702A & 1.16 & 138.2 & 4.63 & $0.97^{+0.01}_{-0.01}$ & $1.45^{+0.02}_{-0.02}$ & $0.94^{+0.06}_{-0.06}$ & \dots & $1.17^{+0.03}_{-0.03}$ & $\dots$ & $0.96^{+0.06}_{-0.06}$ & y &   ISM*/Wind& 4 \\\hline
\bottomrule
\end{tabular}
\par\smallskip
\parbox{\textwidth}{\footnotesize
\textit{References}: (1) \citet{2015ApJ...805...13L};
(2) \citet{2015ApJS..219....9W};
(3) \citet{2023AA...675A.117R};
(4) this work.
}
\end{table*}

In Tables~\ref{MyTabA}--\ref{MyTabD}, the redshift, $T_{\rm 90}$, the break times $T_{\rm b,X}$ ($T_{\rm b,o}$), the temporal indices before the break $\alpha_{\rm X,1}$ ($\alpha_{\rm o,1}$), and the spectral indices during the plateau phase $\beta_{\rm X}$ ($\beta_{\rm o}$) are presented. Wherein, the subscripts ``X'' and ``o'' denote quantities in the X-ray and optical bands, respectively. To distinguish whether the plateaus represent an ``internal plateau'', which is followed by a sharp drop with a temporal decay index of $\alpha > 3$, we also report the temporal indices after the break $\alpha_{\rm X,2}$ and $\alpha_{\rm o,2}$. Here, it is noted that, for GRBs without reported $\beta_{\rm X}$ and $\beta_{\rm o}$, we estimate $\beta_{\rm X}$ from the \emph{Swift} photon index near the break time as $\beta_{\rm X} = \Gamma_{\rm X} - 1$ and adopt $\beta_{\rm o}$ following \citet{2024MNRAS.533.4023D}.

\tabletypesize{\scriptsize}
\begin{table*}
\label{MyTabC}
\caption{Dataset 3: Plateaus detected only in the optical Band}
\centering
\setlength{\tabcolsep}{3pt}
\begin{tabular}{llllllllllllll}
\toprule
\hline
GRB & $z$ & $T_{\rm{90}}$ & $T_{\mathrm{b,X}}$ & $\alpha_{\mathrm{X},1}$ & $\alpha_{\mathrm{X},2}$ & $\beta_{\mathrm{X}}$ & $T_{\mathrm{b,o}}$ & $\alpha_{\mathrm{o},1}$  & $\alpha_{\mathrm{o},2}$  & $\beta_{\mathrm{o}}$ &Radio?&Environment& Ref \\
~&~ & $(\mathrm{s})$ & $(\rm{ks})$ & ~ & ~ & ~ &$(\rm{ks})$  & ~ & ~ & ~ & ~ &~ \\ 
\hline
050820A & 2.61 & 26 & 2379 & $1.12^{+0.08}_{-0.08}$ & $1.89^{+0.11}_{-0.11}$ & $0.89^{+0.05}_{-0.05}$ & ~ & $0.91^{+0.02}_{-0.02}$ & $1.67^{+0.09}_{-0.09}$ & $0.72^{+0.03}_{-0.03}$ & y &Wind& 1  \\ 
051028 & 3.5 & 12 & \dots & $1.16^{+0.08}_{-0.08}$ & \dots & $0.95^{+0.15}_{-0.15}$ & \dots & $0.99^{+0.06}_{-0.06}$ & \dots & $0.60^{+0.00}_{-0.00}$ & &Wind & 1  \\ 
060124 & 2.3 & 710 & \dots & $1.33^{+0.01}_{-0.01}$ & \dots & $0.98^{+0.05}_{-0.05}$ & \dots & $0.88^{+0.02}_{-0.02}$ & \dots & $0.73^{+0.08}_{-0.08}$ & n &Wind& 2  \\ 
060512 & 2.1 & 8.6 & \dots & $1.20^{+0.07}_{-0.07}$ & \dots & $1.04^{+0.10}_{-0.10}$ & \dots & $0.81^{+0.05}_{-0.05}$ & \dots & $0.68^{+0.05}_{-0.05}$ &  &ISM*/Wind& 1  \\ 
060912A & 0.94 & 5 & \dots & $1.07^{+0.02}_{-0.02}$ & \dots & $0.62^{+0.20}_{-0.20}$ & \dots & $0.94^{+0.03}_{-0.03}$ & \dots & $0.60^{+0.15}_{-0.15}$ & n &Wind& 1  \\ 
080804 & 2.20 & 34 & \dots & $1.11^{+0.01}_{-0.01}$ & \dots & $0.82^{+0.10}_{-0.10}$ & \dots & $0.87^{+0.01}_{-0.01}$ & \dots & 0.43 & & & 1  \\ 
081109A & 0.98 & 190 & \dots & $1.22^{+0.02}_{-0.02}$ & \dots & $1.20^{+0.13}_{-0.13}$ & \dots & $0.94^{+0.03}_{-0.03}$ & \dots & 0.40 & n & ISM*/Wind* & 2  \\ 
$090510^{\rm s}$ & 0.90 & 0.3 & 1.5 & $2.27^{+0.06}_{-0.06}$ & \dots & $0.75^{+0.12}_{-0.12}$ & \dots & $0.84^{+0.05}_{-0.05}$ & \dots & $0.85^{+0.05}_{-0.05}$ & n && 1  \\ 
120711A & 1.40 & 44 & \dots & $1.64^{+0.05}_{-0.05}$ & \dots & $0.81^{+0.10}_{-0.10}$ & \dots & $0.96^{+0.04}_{-0.04}$ & \dots & $0.52^{+0.02}_{-0.02}$ &  &Wind*& 1  \\ 
120729A & 0.8 & 71.5 & 6.61 & $1.09^{+0.06}_{-0.06}$ & $2.40^{+0.16}_{-0.16}$ & $0.80^{+0.17}_{-0.17}$ & ~ & $0.94^{+0.05}_{-0.05}$ & $2.27^{+0.09}_{-0.09}$ & $1.00^{+0.10}_{-0.10}$ & n &ISM*/Wind& 1  \\ 
130427A & 0.34 & 138 & \dots & $1.45^{+0.08}_{-0.08}$ & \dots & $0.85^{+0.17}_{-0.17}$ & \dots & $1.00^{+0.02}_{-0.02}$ & \dots & $0.92^{+0.10}_{-0.10}$ &y &Wind & 3  \\ 
141221A & 1.45 & 36.9 & \dots & $1.17^{+0.05}_{-0.05}$ & \dots & $1.09^{+0.31}_{-0.29}$ & $0.20$ & $-0.24^{+0.06}_{-0.06}$ & $1.06^{+0.02}_{-0.02}$ & $0.64^{+0.03}_{-0.03}$ &  & ISM*/Wind& 4  \\ 
180205A & 1.41 & 15.5 & \dots & $1.04^{+0.03}_{-0.03}$ & \dots & $0.81^{+0.21}_{-0.20}$ & \dots & $0.77^{+0.00}_{-0.00}$ & \dots & $0.52^{+0.02}_{-0.02}$ &  &Wind& 4  \\ 
180720B & 0.65 & 48.9 & $51.3$ & $1.16^{+0.09}_{-0.09}$ & $2.16^{+2.48}_{-2.48}$ & $0.76^{+0.04}_{-0.04}$ & \dots & $0.87^{+0.04}_{-0.04}$ & \dots & $0.8^{+0.04}_{-0.04}$ & y &  Wind&  3 \\
190114C & 0.43 & 361 & \dots & $1.50^{+0.02}_{-0.02}$ & \dots & $0.92^{+0.11}_{-0.10}$ & \dots & $0.05^{+0.02}_{-0.02}$ & \dots & $0.83^{+0.04}_{-0.04}$ & y && 4 \\ 
191221B & 1.15 & 8.85 & 90 & $1.18^{+0.04}_{-0.04}$ & $2.3^{+0.2}_{-0.2}$ & $0.89$ & 40 & $0.48^{+0.01}_{-0.01}$ & $1.96^{+0.05}_{-0.05}$ & $0.89$ & y && 5 \\ 
220101A & 4.62 & 162 & 78.7 & $1.15^{+0.02}_{-0.02}$ & $1.74^{+0.05}_{-0.05}$& $0.62^{+0.05}_{-0.05}$ & \dots & $0.88^{+0.02}_{-0.02}$ & \dots & $0.7^{+0.05}_{-0.05}$ & y &Wind*& 4  \\ \hline
\bottomrule
\end{tabular}
\par\smallskip
\parbox{\textwidth}{\footnotesize
\textit{References}: 
(1) \citet{2015ApJS..219....9W};
(2) \citet{2015ApJ...805...13L};
(3)\citet{2025ApJ...978...51D};
(4) this work;
(5) \citet{2024ApJ...972..158C}.
}
\end{table*}

\tabletypesize{\scriptsize}
\begin{table*}
\caption{Dataset 4: No Plateaus Detected}
\centering
\begin{tabular}{llllllllllllll}
\toprule
\hline
GRB & $z$ & $T_{\rm{90}}$ & $T_{\mathrm{b,X}}$ & $\alpha_{\mathrm{X},1}$ & $\alpha_{\mathrm{X},2}$ & $\beta_{\mathrm{X}}$ & $T_{\mathrm{b,o}}$ & $\alpha_{\mathrm{o},1}$  & $\alpha_{\mathrm{o},2}$  & $\beta_{\mathrm{o}}$ & Radio? & Environment & Ref \\
~&~ & $(\mathrm{s})$ & $(\rm{ks})$ & ~ & ~ & ~ &$(\rm{ks})$  & ~ & ~ & ~ & ~&~ &~ \\ 
\hline
050603 & 2.82 & 6 & \dots & $1.71^{+0.05}_{-0.05}$ & \dots & $1.02^{+0.13}_{-0.13}$ & \dots &  $1.70^{+0.13}_{-0.13}$ & \dots &  $0.20^{+0.10}_{-0.10}$ & y& & 1  \\ 
061007 & 1.26 & 75 & \dots & $1.66^{+0.07}_{-0.07}$ & \dots & $1.00^{+0.10}_{-0.10}$ & \dots & $1.62^{+0.08}_{-0.08}$ & \dots & $1.02^{+0.05}_{-0.05}$ & n &Wind*& 1  \\ 
070318 & 0.84  & 63 & \dots & $1.03^{+0.02}_{-0.02}$ & \dots & $0.97^{+0.11}_{-0.11}$ & \dots & $1.02^{+0.10}_{-0.10}$ & \dots & $0.78^{+0.10}_{-0.10}$ &  &ISM*/Wind &1  \\ 
071003 & 1.60 & 150 & \dots & $1.63^{+0.02}_{-0.02}$ & \dots & $0.91^{+0.12}_{-0.12}$ & \dots & $1.62^{+0.11}_{-0.11}$ & \dots & $0.35^{+0.11}_{-0.11}$ & y & &1  \\ 
080810 & 3.35 & 106 & 17.2 & $1.17^{+0.14}_{-0.14}$ & $1.87^{+0.11}_{-0.11}$ & $1.06^{+0.12}_{-0.12}$ & \dots & $1.20^{+0.01}_{-0.01}$ & \dots & 0.44 & y && 2  \\ 
081203A & 2.3 & 294 & 7.1 & $1.04^{+0.09}_{-0.09}$ & $1.89^{+0.11}_{-0.11}$ & $1.04^{+0.10}_{-0.10}$ & ~ & $1.15^{+0.07}_{-0.07}$ & $1.87^{+0.13}_{-0.13}$ & 0.60 & n &Wind*& 1  \\ 
090323 & 3.57 & 150 & \dots & $1.62^{+0.09}_{-0.09}$ & \dots & $0.87^{+0.22}_{-0.22}$ & \dots & $1.55^{+0.05}_{-0.05}$ & \dots & $0.74^{+0.15}_{-0.15}$ & y &Wind*& 1  \\ 
090328 & 0.74 & 80 & \dots & $1.67^{+0.11}_{-0.11}$ & \dots & $0.90^{+0.30}_{-0.30}$ & \dots & $1.84^{+0.08}_{-0.08}$ & \dots & $1.19^{+0.21}_{-0.21}$ & y &Wind &1  \\ 
090812 & 2.45 & 66.7 & \dots & $1.22^{+0.09}_{-0.09}$ & \dots & $0.89^{+0.14}_{-0.14}$ & \dots & $1.27^{+0.05}_{-0.05}$ & \dots & 0.36 & n && 1  \\ 
090902B & 1.82 & 21 & \dots & $1.40^{+0.03}_{-0.03}$ & \dots & $0.82^{+0.12}_{-0.12}$ & \dots & $1.00^{+0.06}_{-0.06}$ & \dots & $0.68^{+0.11}_{-0.11}$ & y & Wind&2  \\ 
100728B & 2.11 & 12.1 & \dots & $1.51^{+0.07}_{-0.07}$ & \dots & $1.08^{+0.18}_{-0.18}$ & \dots & $1.01^{+0.02}_{-0.02}$ & \dots & $0.7^{+0.18}_{-0.18}$ &  &ISM*/Wind& 2  \\ 
110205A & 2.22 & 257 & \dots & $1.59^{+0.02}_{-0.02}$ & \dots & $0.78^{+0.06}_{-0.06}$ & \dots & $1.51^{+0.08}_{-0.08}$ & \dots & $0.49^{+0.08}_{-0.08}$ & y && 1  \\ 
110918A & 0.98 & 22 & \dots & $1.61^{+0.12}_{-0.12}$ & \dots & $0.89^{+0.30}_{-0.30}$ & \dots & $1.65^{+0.07}_{-0.07}$ & \dots & $0.42^{+0.18}_{-0.18}$ &  & &1  \\ 
141220A & 1.32 & 7.21 & \dots & $1.38^{+0.06}_{-0.06}$ & \dots & $0.77^{+0.29}_{-0.28}$ & \dots & $1.10^{+0.01}_{-0.01}$ & \dots & $0.6^{+0.2}_{-0.2}$ &  &Wind& 3  \\ 
170405A & 3.51 & 164.7 & \dots & $1.76^{+0.07}_{-0.07}$ & \dots & $0.97^{+0.14}_{-0.11}$ & \dots & $1.36^{+0.10}_{-0.10}$ & \dots & $0.8^{+0.09}_{-0.09}$ &  & Wind* & 3  \\
210822A & 1.74 & 180 & 11.5 & $1.03^{+0.09}_{-0.09}$ & $1.82^{+0.05}_{-0.05}$ & $0.74^{+0.02}_{-0.02}$ & \dots & $1.34^{+0.02}_{-0.02}$ & \dots & $0.77^{+0.03}_{-0.03}$ & y  & Wind*  & 4  \\
221009A & 0.15 & 1105 & \dots & $1.13^{+0.01}_{-0.01}$ & $1.82^{+0.07}_{-0.07}$ & $0.77^{+0.20}_{-0.19 }$ & \dots & $1.07^{+0.01}_{-0.01}$ & \dots & $0.75^{+0.02}_{-0.02}$ & y &Wind*& 3  \\ \hline
\bottomrule
\end{tabular}
\par\smallskip
\parbox{\textwidth}{\footnotesize
\textit{References}: 
(1) \citet{2015ApJS..219....9W};
(2) \citet{2015ApJ...805...13L};
(3) this work
(4)\citet{2025ApJ...978...51D}.
}
\label{MyTabD}
\end{table*}

\begin{table*}
\centering
\caption{Closure relations under the energy-injection model for a relativistic, isotropic, self-similar deceleration phase with $\nu_{\rm{a}} < \rm min (\nu_{\rm{m}},\nu_{\rm{c}})$, considering both ISM and wind cases, as well as fast-cooling (FC) and slow-cooling (SC) regimes \citep{2013NewAR..57..141G}.}
\label{MyTabE}
\resizebox{0.8\textwidth}{!}{
\begin{tabular}{lcccc}
\toprule
\hline
& $\nu$ & $\beta$ & $\alpha(\beta) (q>2)$ & $\alpha(\beta) (1<q<2)$ \\
\midrule
ISM, SC & $\nu_{\mathrm{m}}<\nu<\nu_{\mathrm{c}}$ & $\dfrac{p-1}{2}$ & $(q-1)+\dfrac{(2+q)\beta}{2}$ & $\dfrac{19q-10}{16}+\dfrac{(2+q)\beta}{8}$ \\
& $\nu>\nu_{\mathrm{c}}$ & $\dfrac{p}{2}$ & $\dfrac{(q-2)}{2}+\dfrac{(2+q)\beta}{2}$ & $\dfrac{7q-2}{8}+\dfrac{(2+q)\beta}{8}$ \\
ISM, FC & $\nu>\nu_{\mathrm{m}}$ & $\dfrac{p}{2}$ & $\dfrac{(q-2)}{2}+\dfrac{(2+q)\beta}{2}$ & $\dfrac{7q-2}{8}+\dfrac{(2+q)\beta}{8}$ \\
Wind, SC & $\nu_{\mathrm{m}}<\nu<\nu_{\mathrm{c}}$ & $\dfrac{p-1}{2}$ & $\dfrac{q}{2}+\dfrac{(2+q)\beta}{2}$ & $\dfrac{5q+4}{8}+\dfrac{\beta q}{4}$ \\
& $\nu>\nu_{\mathrm{c}}$ & $\dfrac{p}{2}$ & $\dfrac{(q-2)}{2}+\dfrac{(2+q)\beta}{2}$ & $\dfrac{(\beta+3)q}{4}$ \\
Wind, FC & $\nu>\nu_{\mathrm{m}}$ & $\dfrac{p}{2}$ & $\dfrac{(q-2)}{2}+\dfrac{(2+q)\beta}{2}$ &  $\dfrac{(\beta+3)q}{4}$ \\
\hline
\bottomrule
\end{tabular}
} 
\end{table*} 

\subsection{Closure Relations with Energy Injections}

GRB afterglows are produced during the interaction between relativistic jets and the circumburst medium in the standard external shock model.  In this framework, the so-called ``closure relations'' are predicted, i.e., under the convention $F_\nu \propto t^{-\alpha}\nu^{-\beta}$, there exists a relationship between the temporal decay index $\alpha$ and the spectral index $\beta$ of GRB afterglows \citep[e.g.,][]{2004IJMPA..19.2385Z,2006ApJ...642..354Z,2013NewAR..57..141G}. The closure relations have been widely applied in the investigations of multi-wavelength afterglows \citep[e.g.,][]{2021PASJ...73..970D,2022ApJ...940..169D}.

The energy-injection model generally involves a long-lasting central engine, whose activity can continuously inject energy into the fireball. The luminosity of the long-lasting central engine can be described as \citep[e.g.,][]{2001ApJ...552L..35Z,2015ApJS..219....9W}
\begin{equation}
L(t) = L_{0}\left(\tfrac{t}{t_{0}}\right)^{-q},
\label{L_{t}}
\end{equation}
where $t_{0}$ is the characteristic timescale roughly representing the external shock start to decelerate, and $L_{0}$ is the characteristic luminosity at the time. $q$ is the luminosity injection index, which is a function of time. However, it is generally assumed as a discrete value for simplicity. Energy injection occurs when the injection parameter $q < 1$, and $q=0$ corresponds to a constant luminosity injection. In this work, we consider two representative regions: $q\in (0, 0.5)$ and $q\in (0, 0.8)$. In our diagrams, we depict the interval $q\in (0, 0.5)$ with darker areas while $q\in (0.5, 0.8)$ is marked separately with lighter areas to illustrate the effect of higher values. The total shaded area encompasses the full region $q\in (0, 0.8)$. We then substituted both cases into the closure relations with the energy injection taken into consideration, which are summarized in Table~\ref{MyTabE}. Here, we adopt a relativistic, isotropic, and self-similar deceleration phase with $\nu_{\rm a} < {\rm min}(\nu_{\rm m}, \nu_{\rm c})$ and ${\rm min}(\nu_{\rm X}, \nu_{\rm o}) > \nu_{\rm m}$, where $\nu_{\rm a}$ is the self-absorption frequency, $\nu_{\rm m}$ is the minimum injection frequency, and $\nu_{\rm c}$ is the cooling frequency. These conditions are commonly satisfied for the X-ray and optical afterglows for typical GRBs. The table covers all relevant spectral regimes, including both interstellar medium (ISM) and wind environments, as well as fast and slow cooling processes.

\begin{figure*}
\centering
\includegraphics[trim=0 0 10 10, angle=0,width=0.24\textwidth]{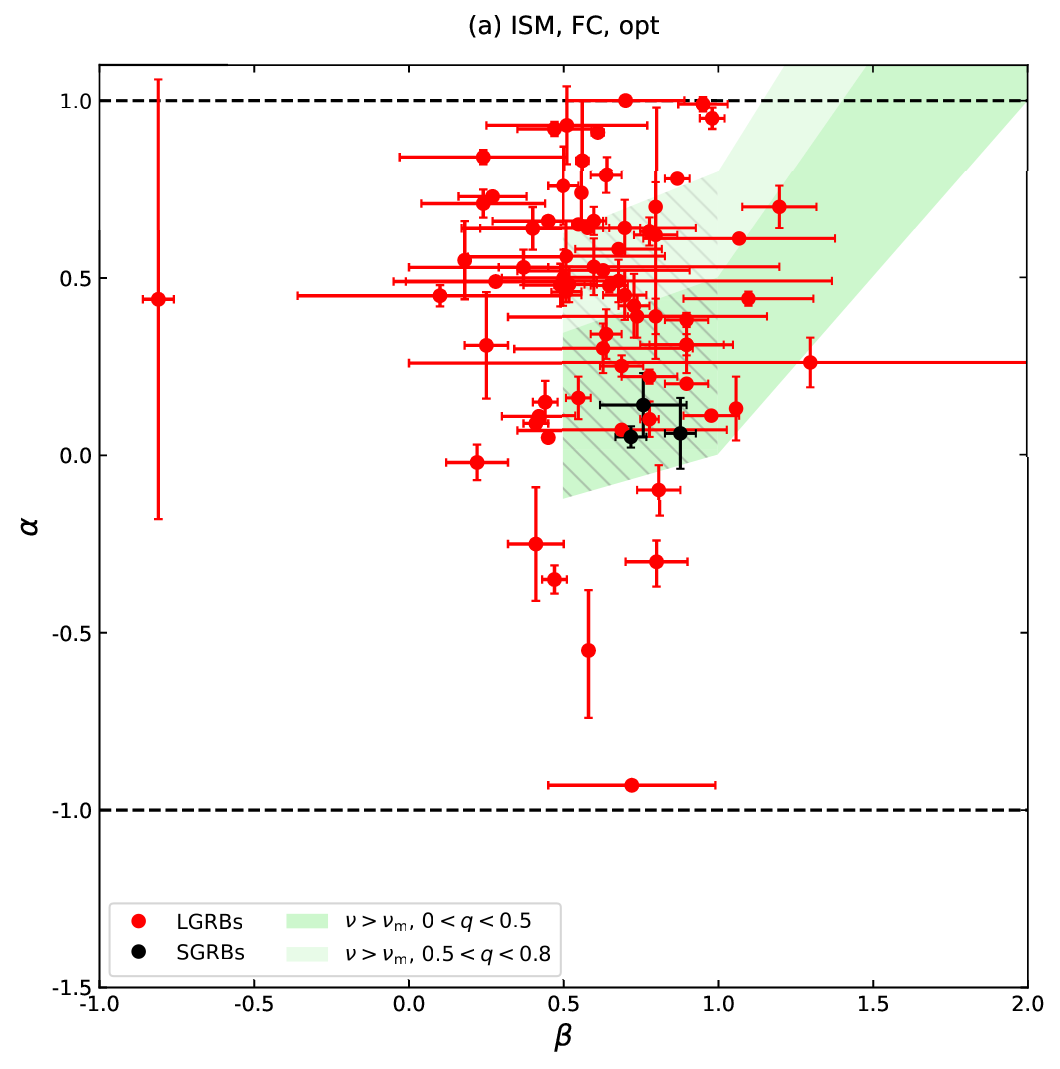}
\includegraphics[trim=0 0 10 10, angle=0,width=0.24\textwidth]{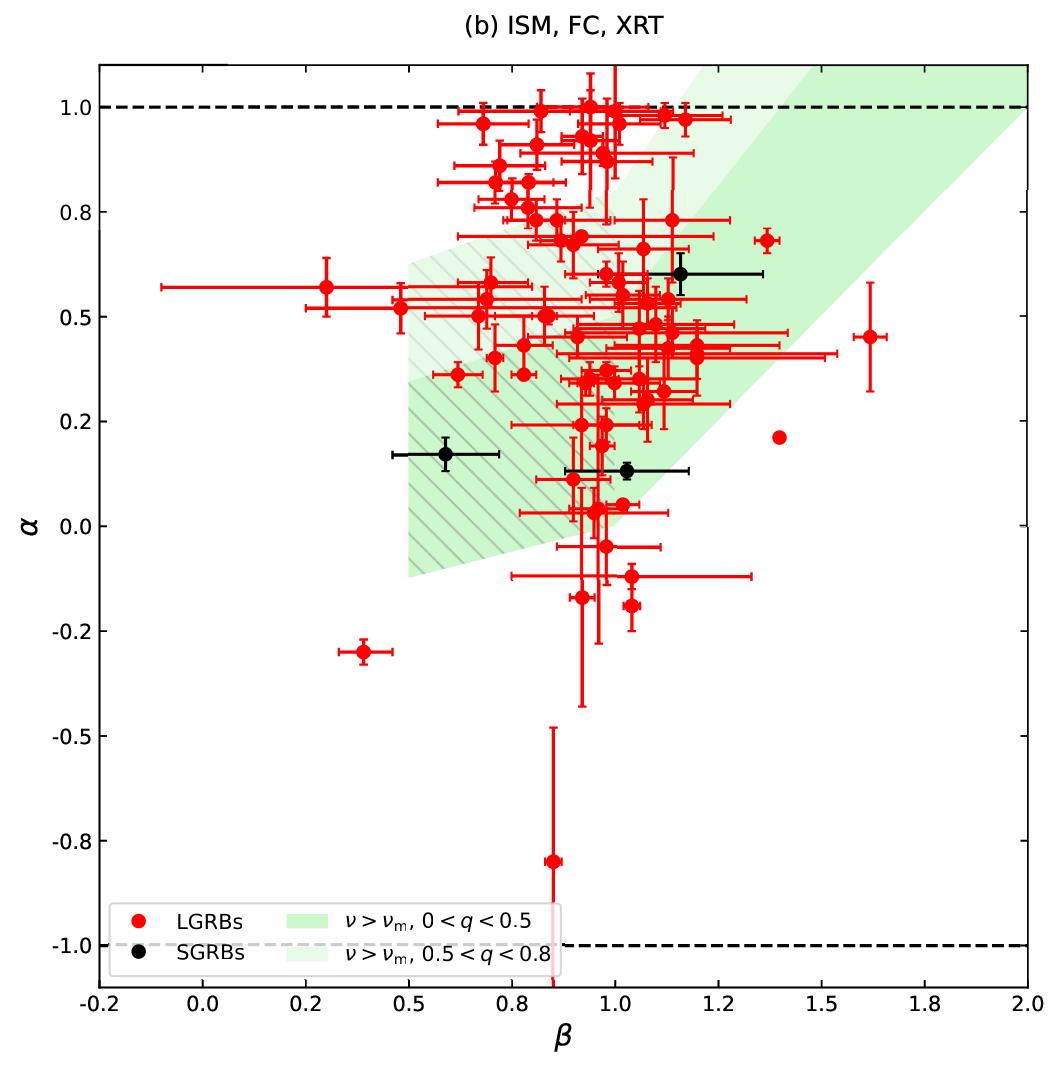}
\includegraphics[trim=0 0 10 10, angle=0,width=0.24\textwidth]{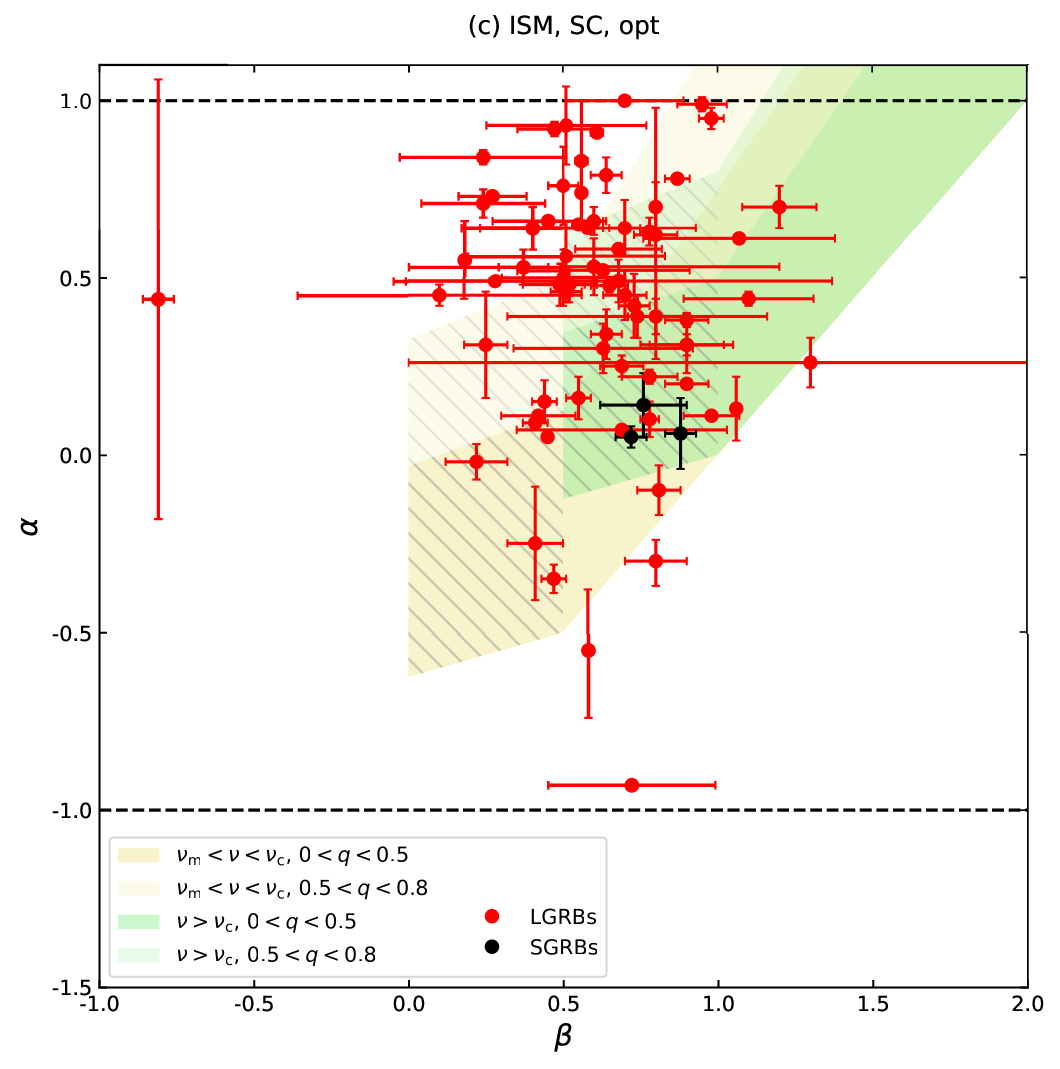}
\includegraphics[trim=0 0 10 10, angle=0,width=0.24\textwidth]{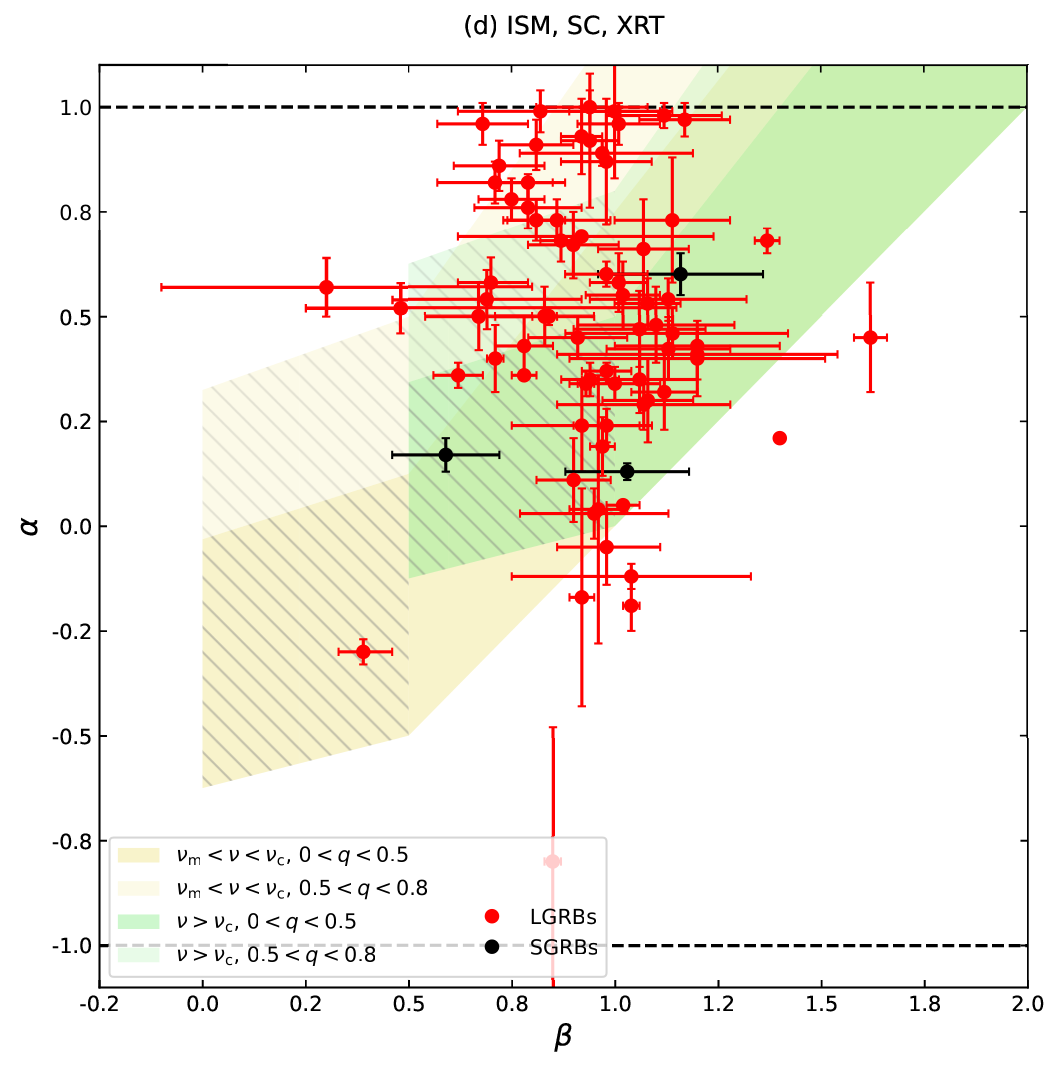}\\
\includegraphics[trim=0 0 10 10, angle=0,width=0.24\textwidth]{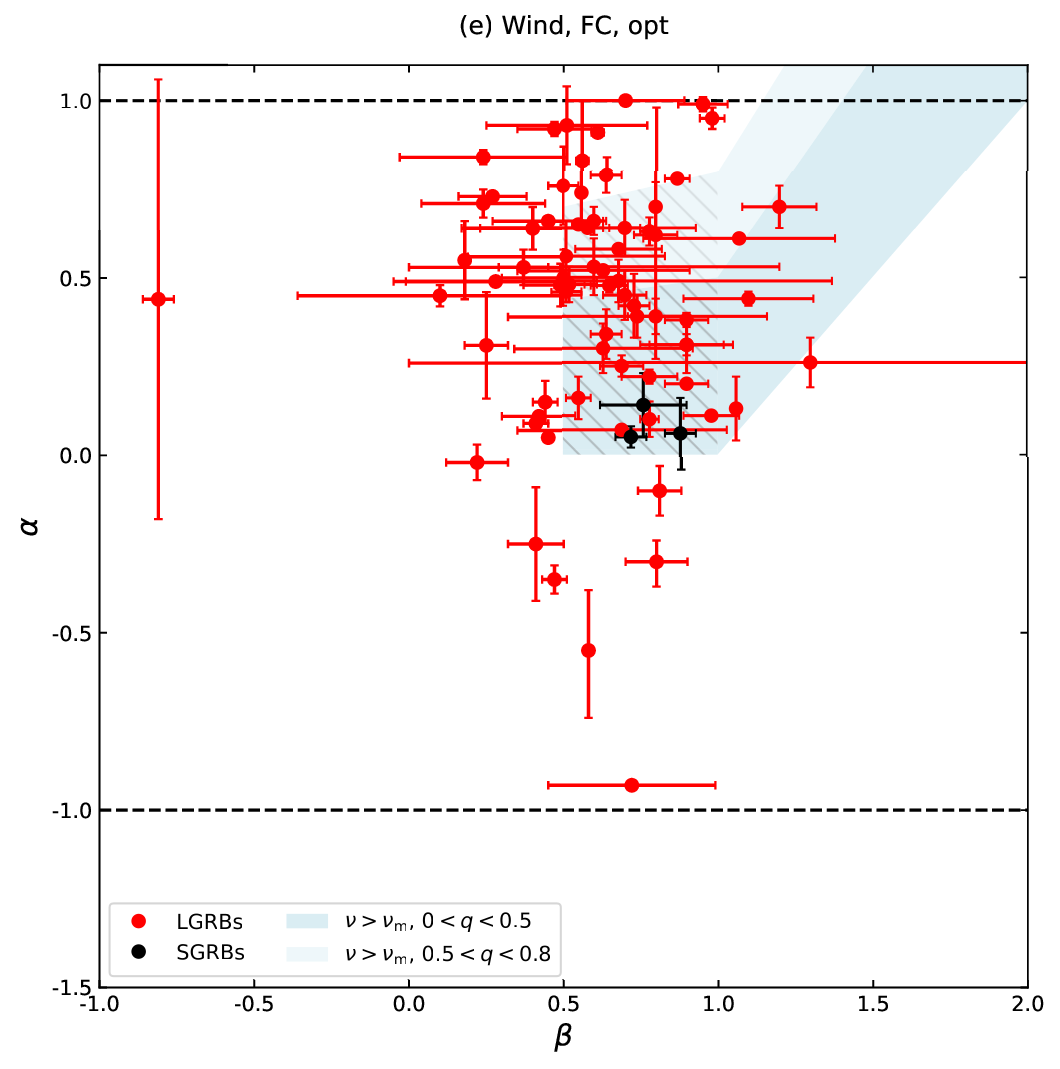}
\includegraphics[trim=0 0 10 10, angle=0,width=0.24\textwidth]{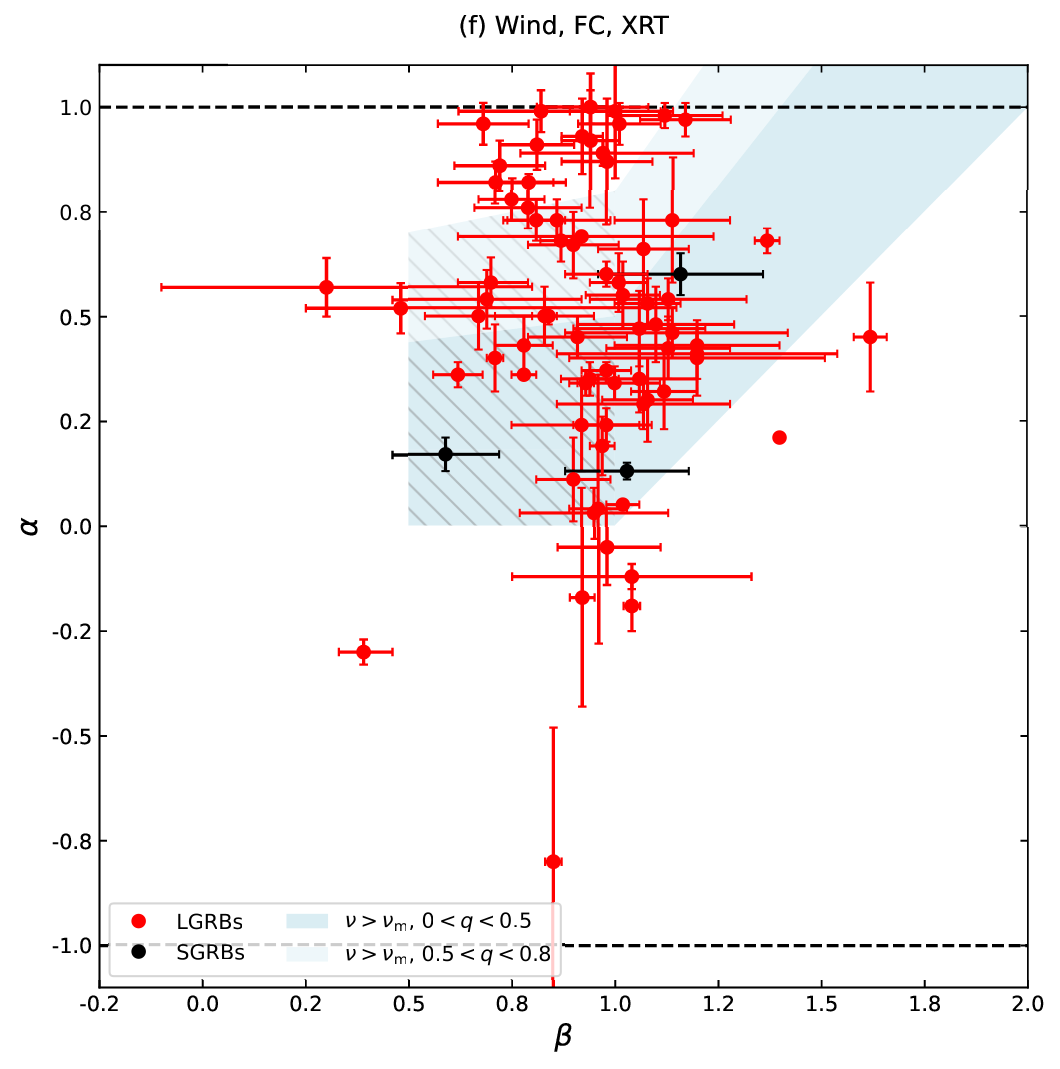}
\includegraphics[trim=0 0 10 10, angle=0,width=0.24\textwidth]{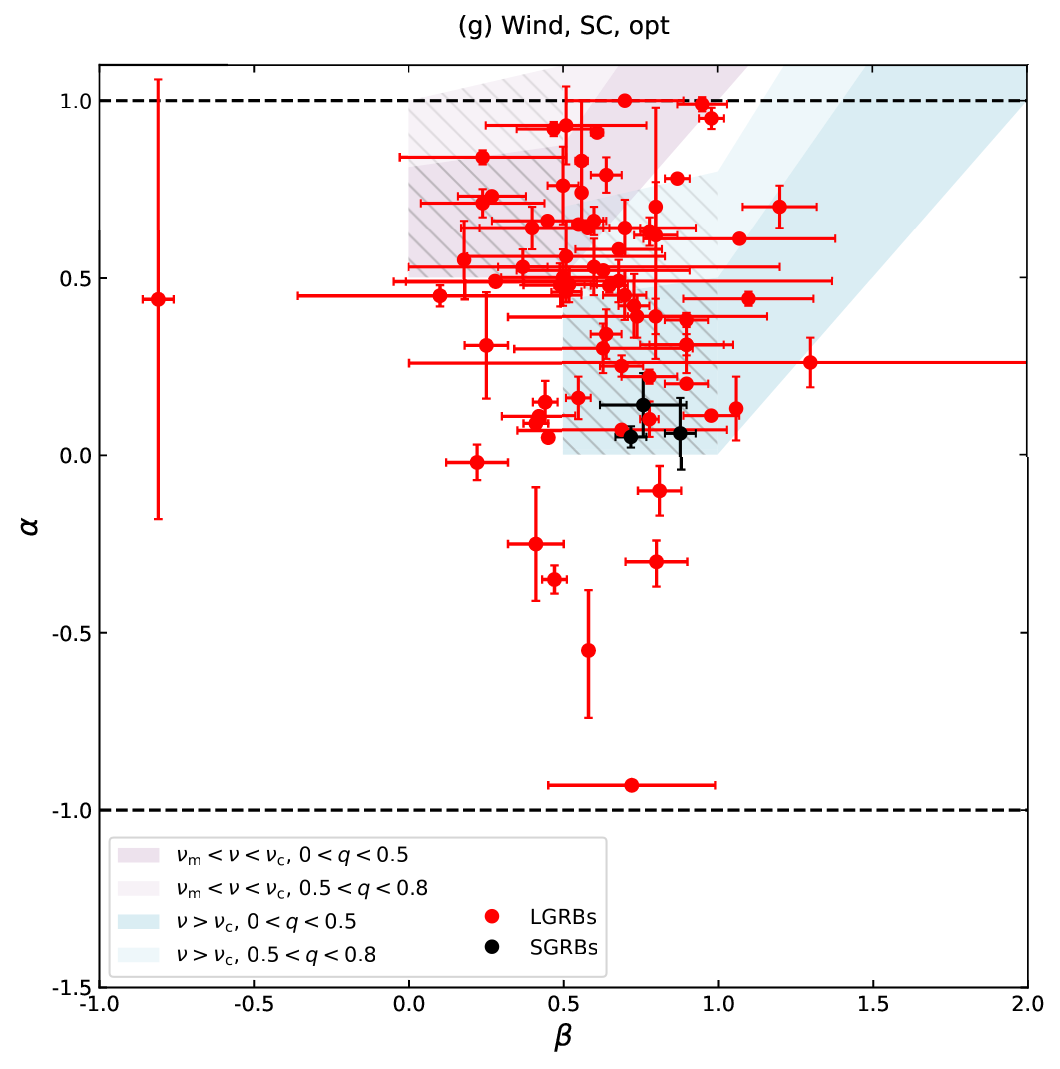}
\includegraphics[trim=0 0 10 10, angle=0,width=0.24\textwidth]{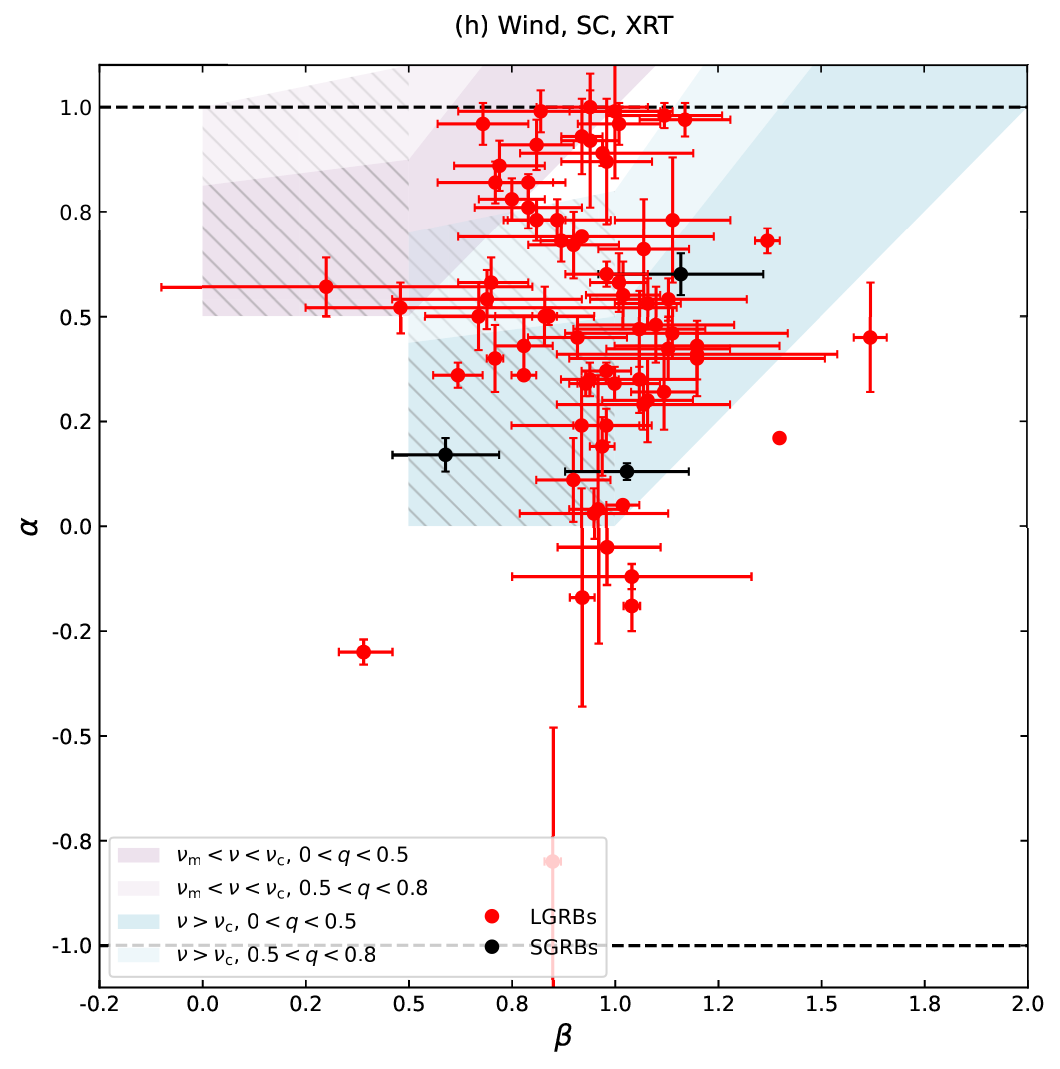}
\caption{Closure relations for dataset~1 with multi-band plateaus. The horizontal black dashed lines indicate $\alpha=\pm1$. The shaded regions correspond to different ranges of the injection parameter $q$: darker colors for $q \in (0,0.5)$ and lighter colors for $q \in (0.5,0.8)$, the combined shaded area represents the full region $q \in (0,0.8)$. Hatched regions denote the case where the electron spectral index lies in the range $1<p<2$. Panels (a)–(h) show the closure relations for different external shock scenarios: (a) ISM, fast cooling, optical band; (b) ISM, fast cooling, X-ray band; (c) ISM, slow cooling, optical band; (d) ISM, slow cooling, X-ray band; (e) wind, fast cooling, optical band; (f) wind, fast cooling, X-ray band; (g) wind, slow cooling, optical band; and (h) wind, slow cooling, X-ray band.
\label{MyFig1}}
\end{figure*}

\subsection{Isotropic Energy of X-ray Plateaus}

Two types of GRB central engines have been proposed, namely a rotating BH surrounded by a hyper-accreting disk \citep[e.g.,][]{1991AcA....41..257P,1999ApJ...524..262M,2017NewAR..79....1L} and a magnetar \citep[e.g.,][]{1992ApJ...392L...9D,1992Natur.357..472U,1998PhRvL..81.4301D,1998AA...333L..87D,2011MNRAS.413.2031M}. For the latter scenario, the energy reservoir is the total rotational energy of the magnetar, $E_{\rm rot}\simeq 2\times10^{52}\,\mathrm{erg}$, whereas the BH model has no such limit \citep[e.g.,][]{2014ApJ...785...74L,2018ApJS..236...26L}. If the central engine is a magnetar, its rotational energy is dissipated mainly through magnetic dipole radiation. Generally, the luminosity $L_{\rm EM}$ of magnetic dipole radiation is injected into the afterglows, leading to plateaus appearing in the X-ray light curves. Thus, the X-ray plateau luminosity can be expressed as \citep[e.g.,][]{2017ApJ...843L...1L,2019ApJ...871...54L}
\begin{equation}
L_{\rm X}=\eta_{\rm X} L_{\rm EM},
\label{eq:L_X}
\end{equation}
where $\eta_{\rm X}$ is the X-ray radiation efficiency and usually assumed to be a constant. Here, we adopt $\eta_{\rm X}=0.1$ following \citet{2019ApJ...871...54L} for a simple estimation. Accordingly, Equation~(\ref{eq:L_X}) leads to a constrain $E_{\rm X}<\eta_{\rm X}E_{\rm rot}$, with $E_{\rm X}$ being the energy of the X-ray plateaus. Under the assumption of quasi-isotropic energy injection, the observed X-ray plateau energy, $E_{\rm X,iso}$, can be obtained by integrating the observed luminosity over time, i.e.,
\begin{equation} \label{E_Xiso}
E_{\rm X,iso} = \int_{t_{\rm s}}^{t_{\rm end}} \frac{4\pi d_{L}^{2}kF(t)}{1+z}dt,
\end{equation}
where $d_{L}$ is the luminosity distance, $k = (1+z)^{\beta - 1}$ is the $k$-correction factor, and $F(t)$ is the observed X-ray flux, $t_{\rm s}$ and $t_{\rm end}$ denote the start and end times of the plateau phase, respectively. According to \citet{2015ApJS..219....9W}, we set $t_{\rm s}=\max\,(T_{90}, 60\,\mathrm{s})$ and $t_{\rm end}$ to be the break time of the plateaus. The duration of the luminosity injection can be roughly written as $t_{\rm inj}=t_{\rm end}-t_{\rm s}$. It should be noted that the luminosity injection of the magnetic dipole radiation directly into X-ray afterglows differs from the energy-injection model (mentioned in Section 2.2) in which energy is injected into the external shocks and multi-band afterglows are produced via the synchrotron radiation of electrons accelerated by the shocks.

\section{results}

\subsection{ The Applicability of the Energy Injection Model}

\begin{figure*}
\centering
\includegraphics[trim=0 0 10 10, angle=0,width=0.24\textwidth]{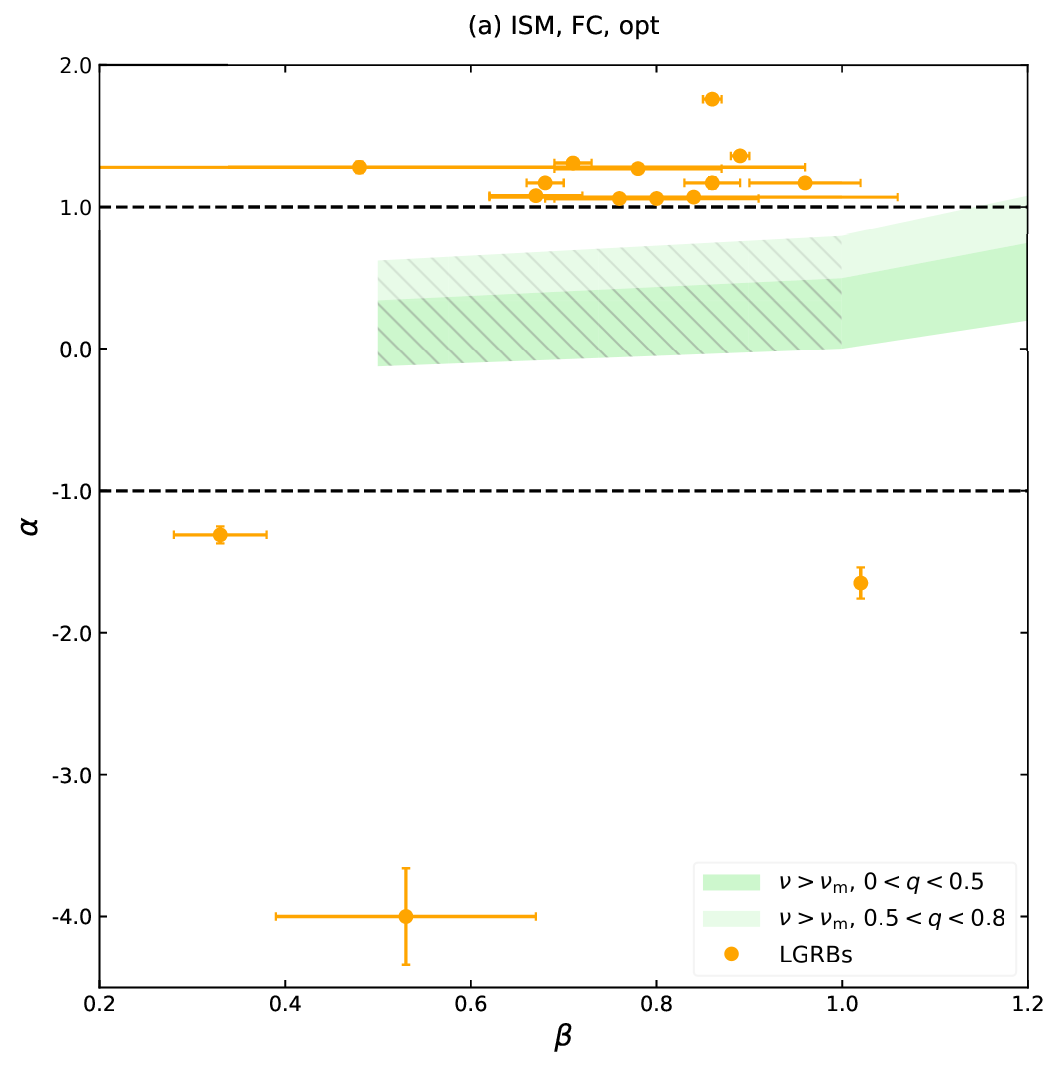}
\includegraphics[trim=0 0 10 10, angle=0,width=0.24\textwidth]{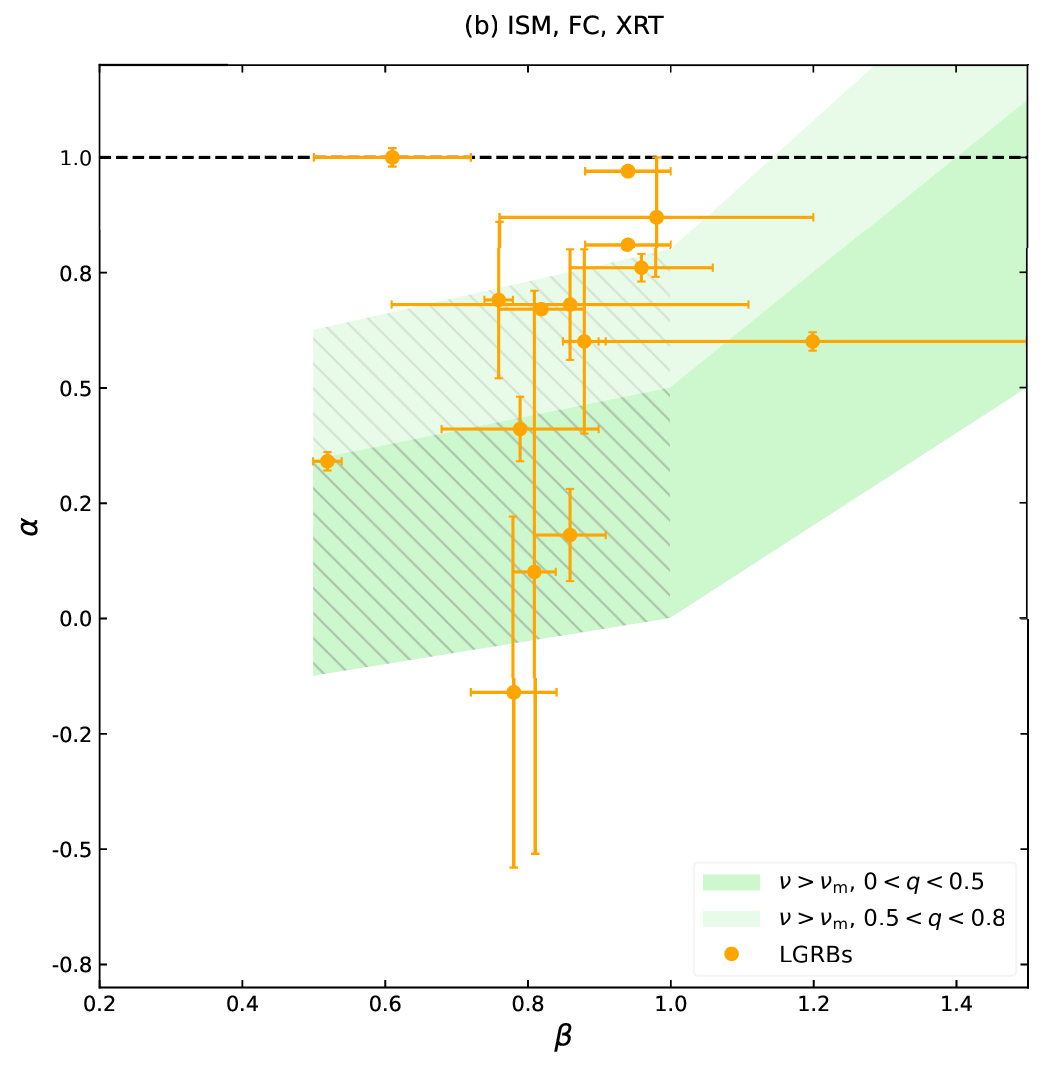}
\includegraphics[trim=0 0 10 10, angle=0,width=0.24\textwidth]{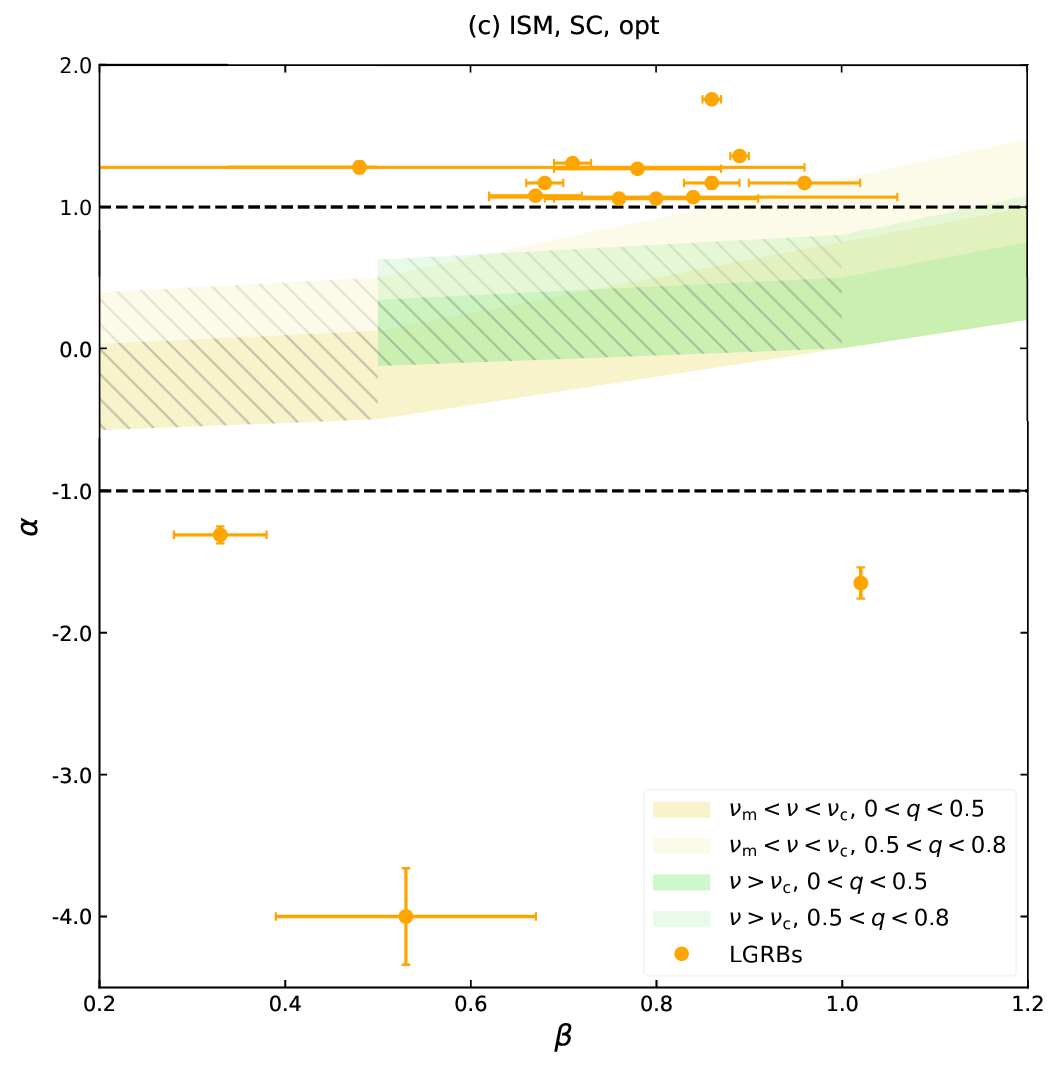}
\includegraphics[trim=0 0 10 10, angle=0,width=0.24\textwidth]{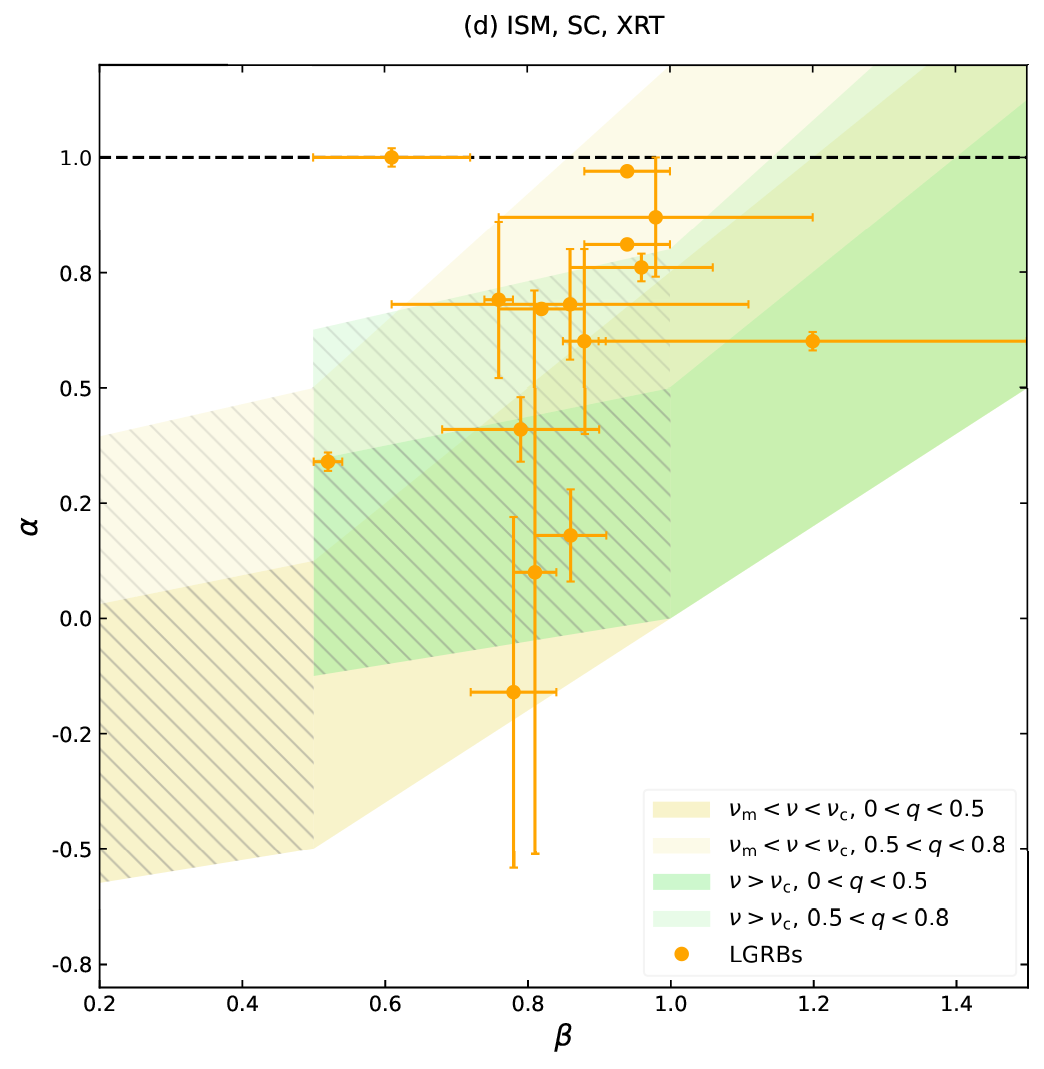}\\
\includegraphics[trim=0 0 10 10, angle=0,width=0.24\textwidth]{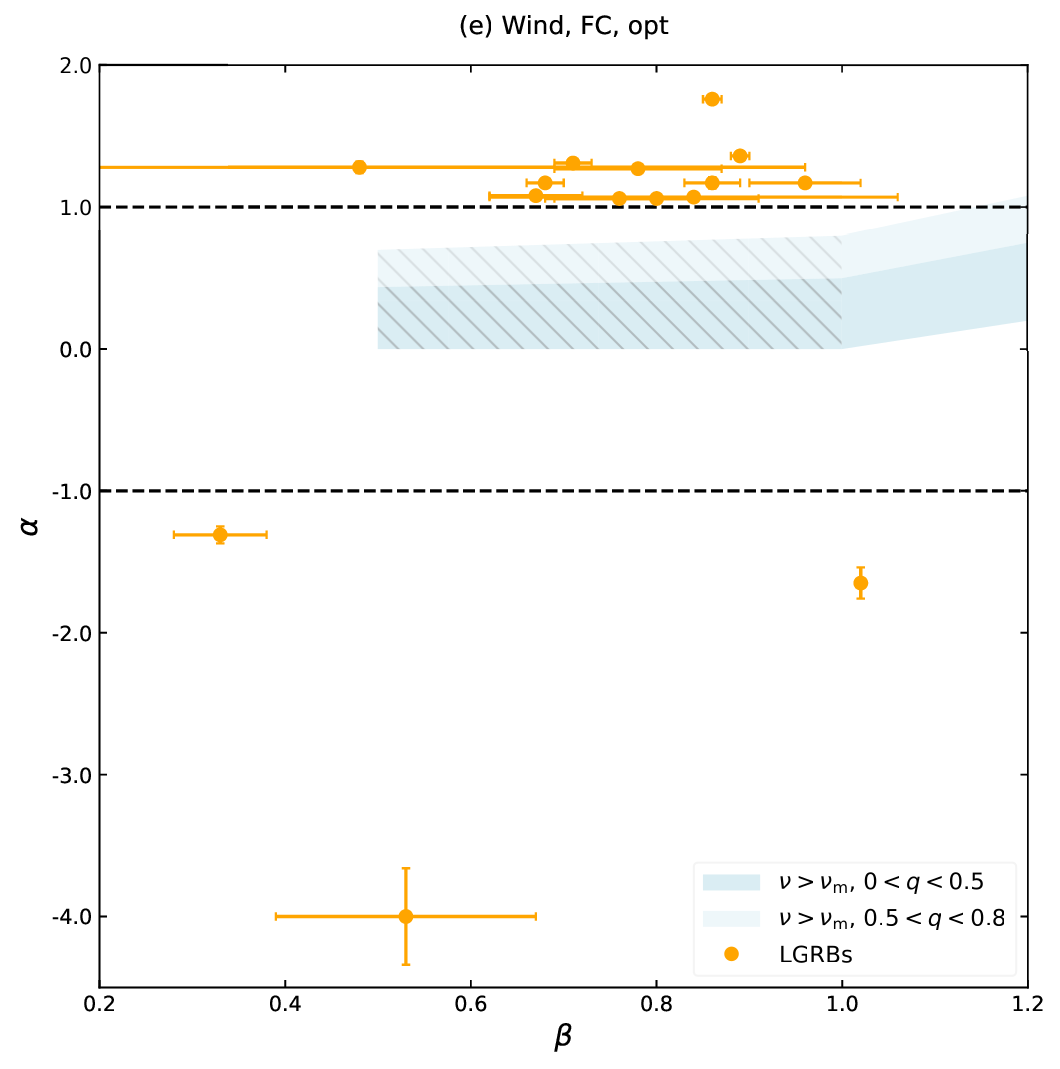}
\includegraphics[trim=0 0 10 10, angle=0,width=0.24\textwidth]{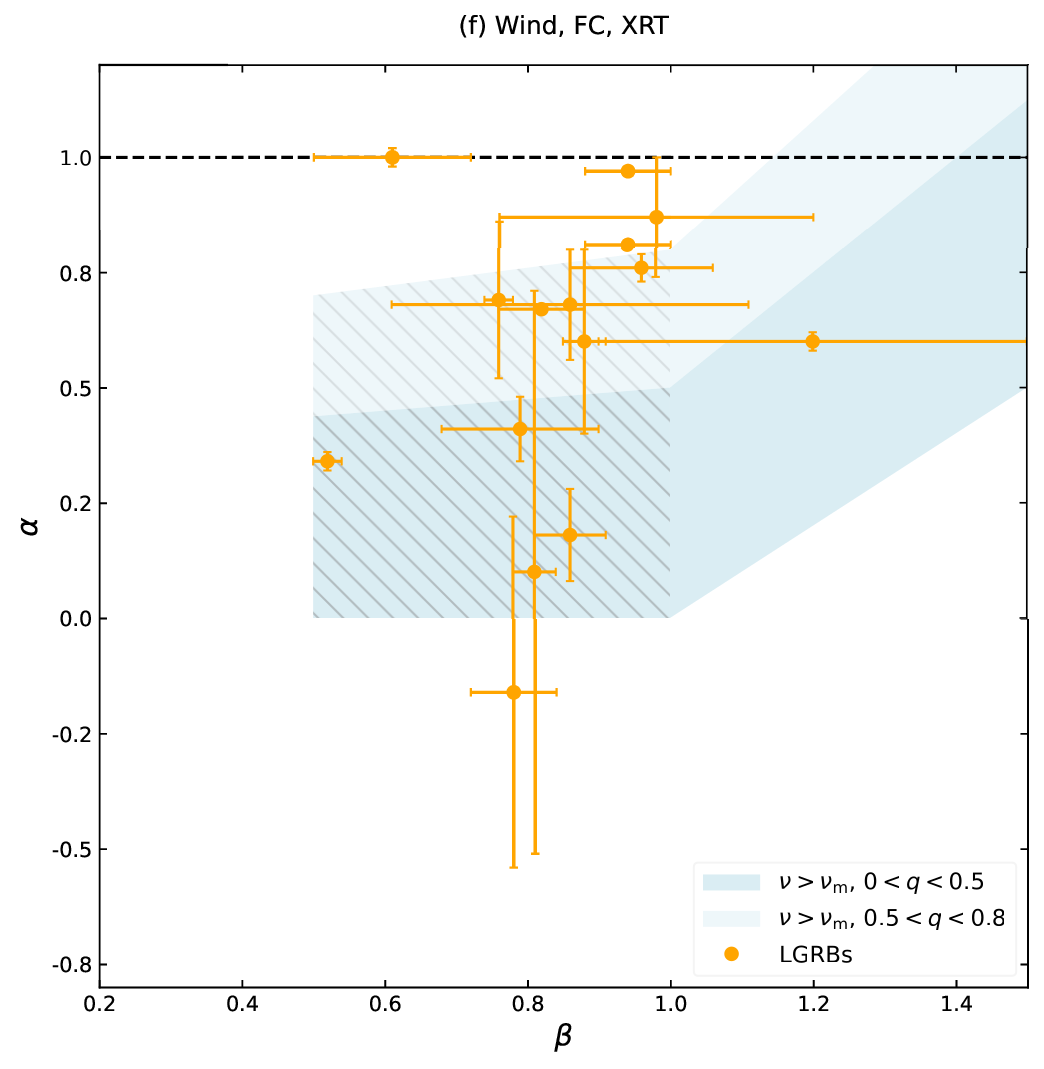}
\includegraphics[trim=0 0 10 10, angle=0,width=0.24\textwidth]{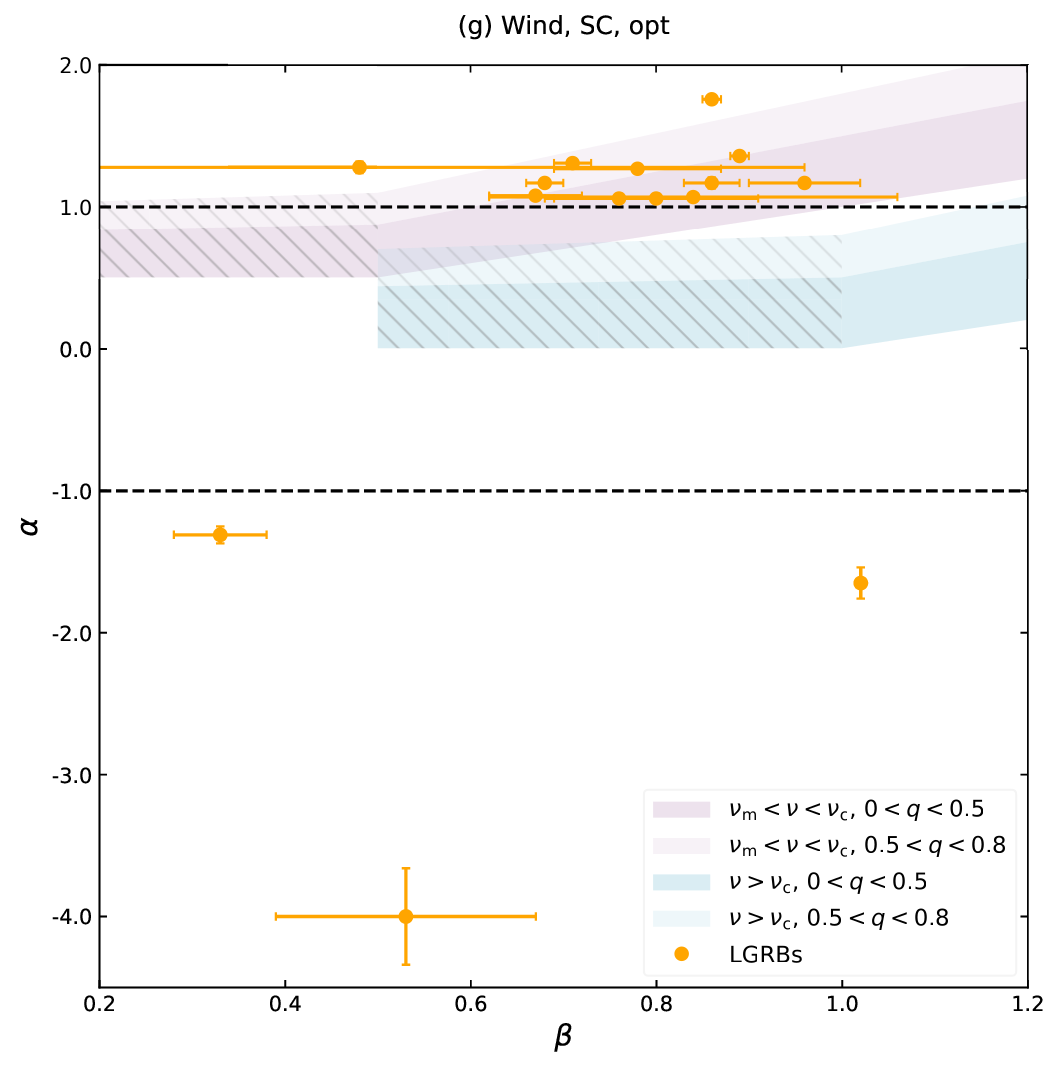}
\includegraphics[trim=0 0 10 10, angle=0,width=0.24\textwidth]{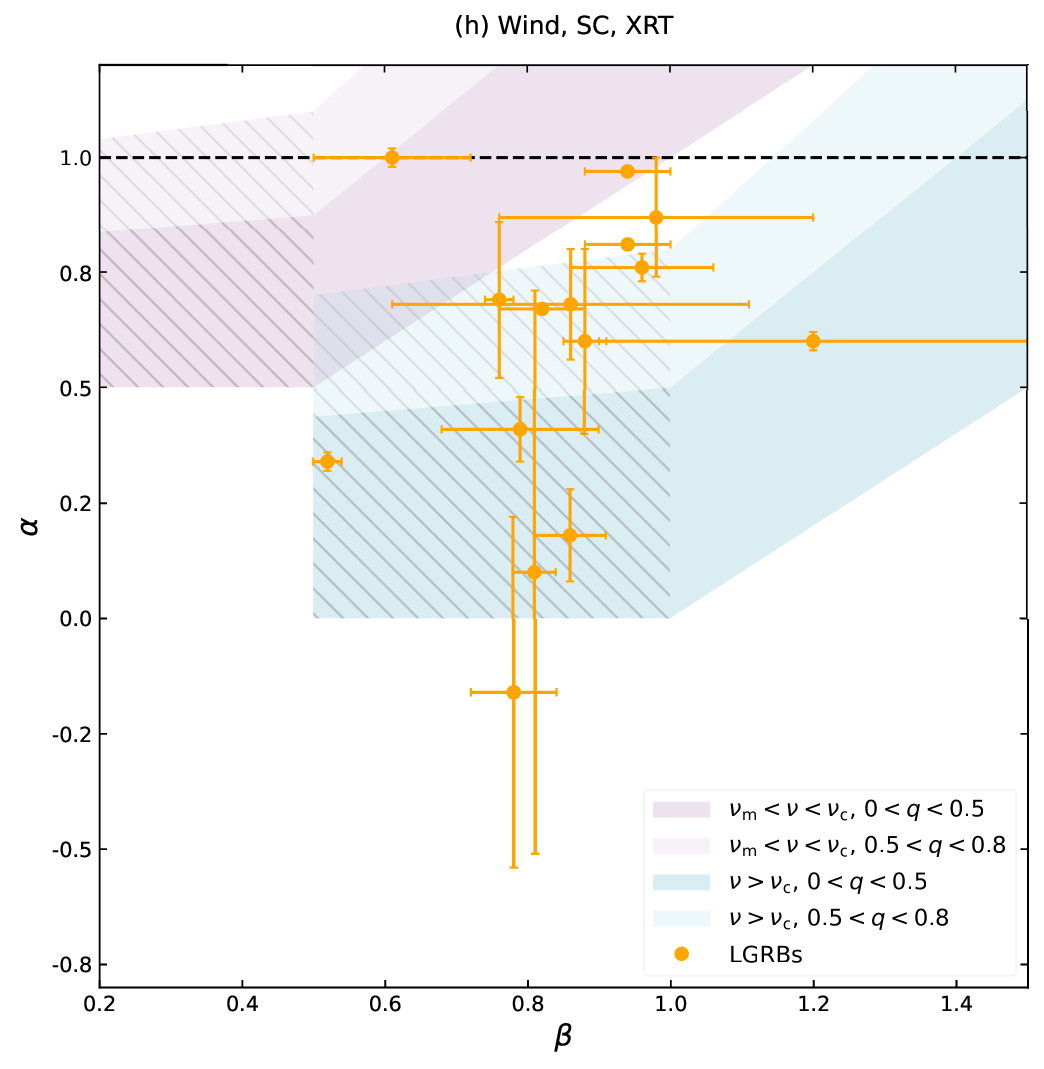}
\caption{Closure relations for dataset~2 with plateaus detected only in the X-ray band, in the same format as dataset~1.
\label{MyFig2}}
\end{figure*}

\begin{figure*}
\centering
\includegraphics[trim=0 0 10 10, angle=0,width=0.24\textwidth]{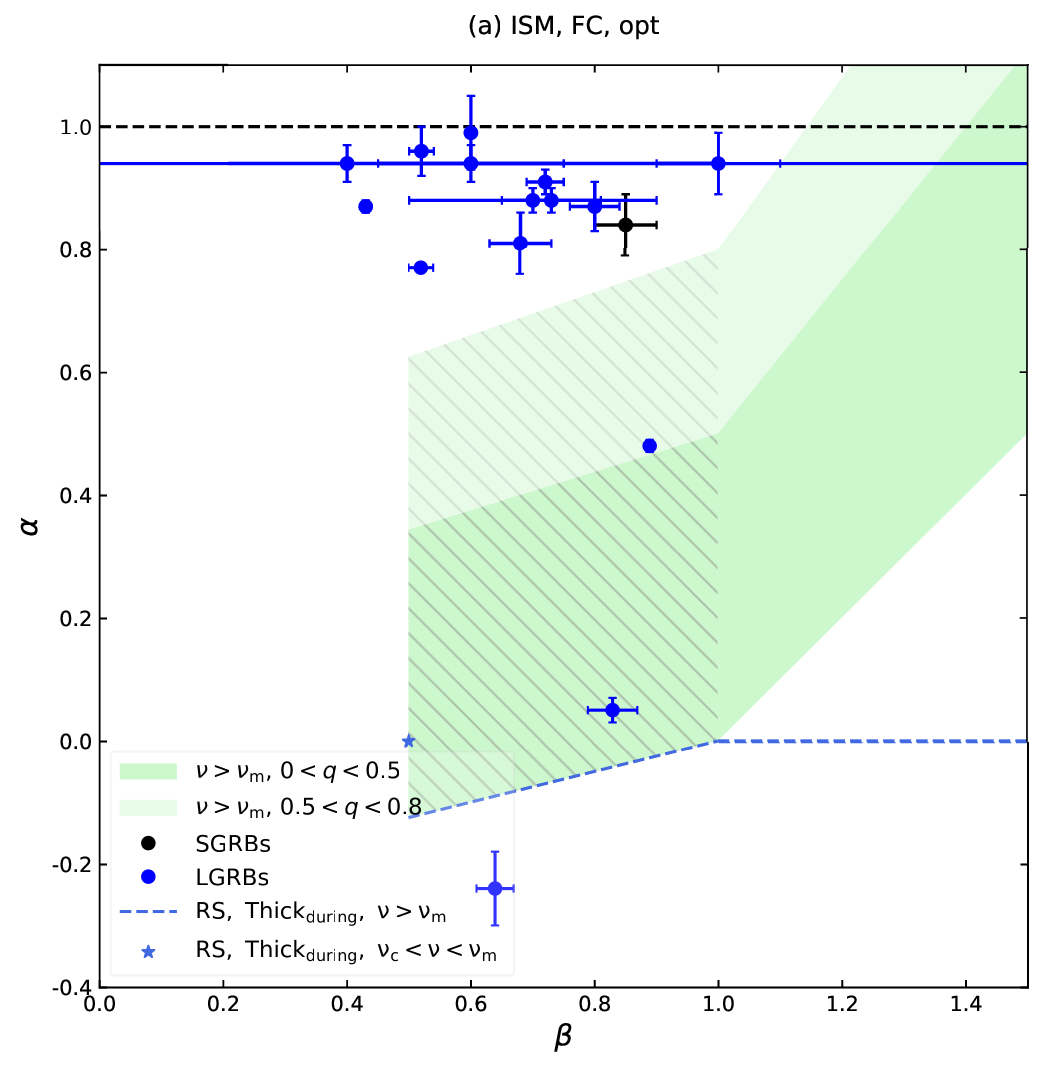}
\includegraphics[trim=0 0 10 10, angle=0,width=0.24\textwidth]{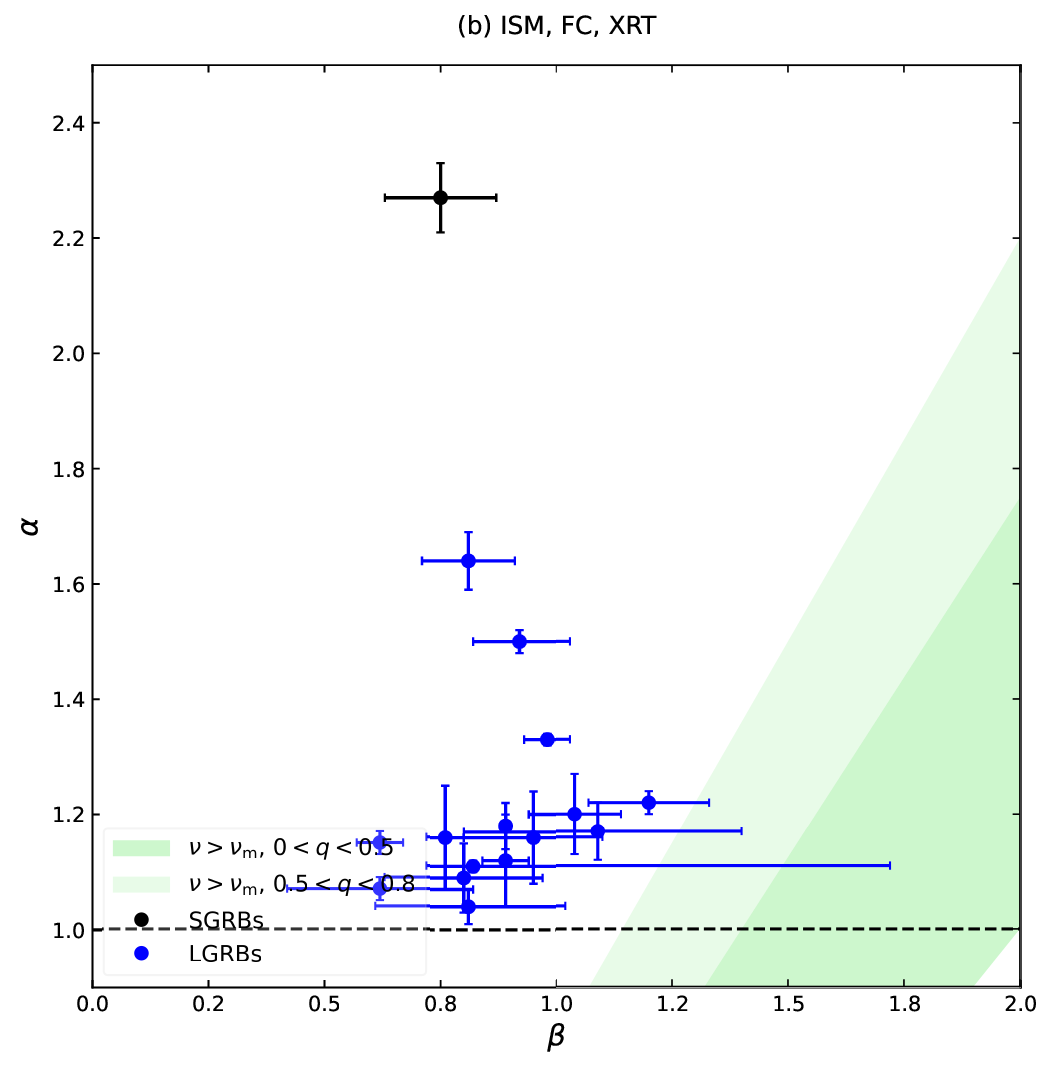}
\includegraphics[trim=0 0 10 10, angle=0,width=0.24\textwidth]{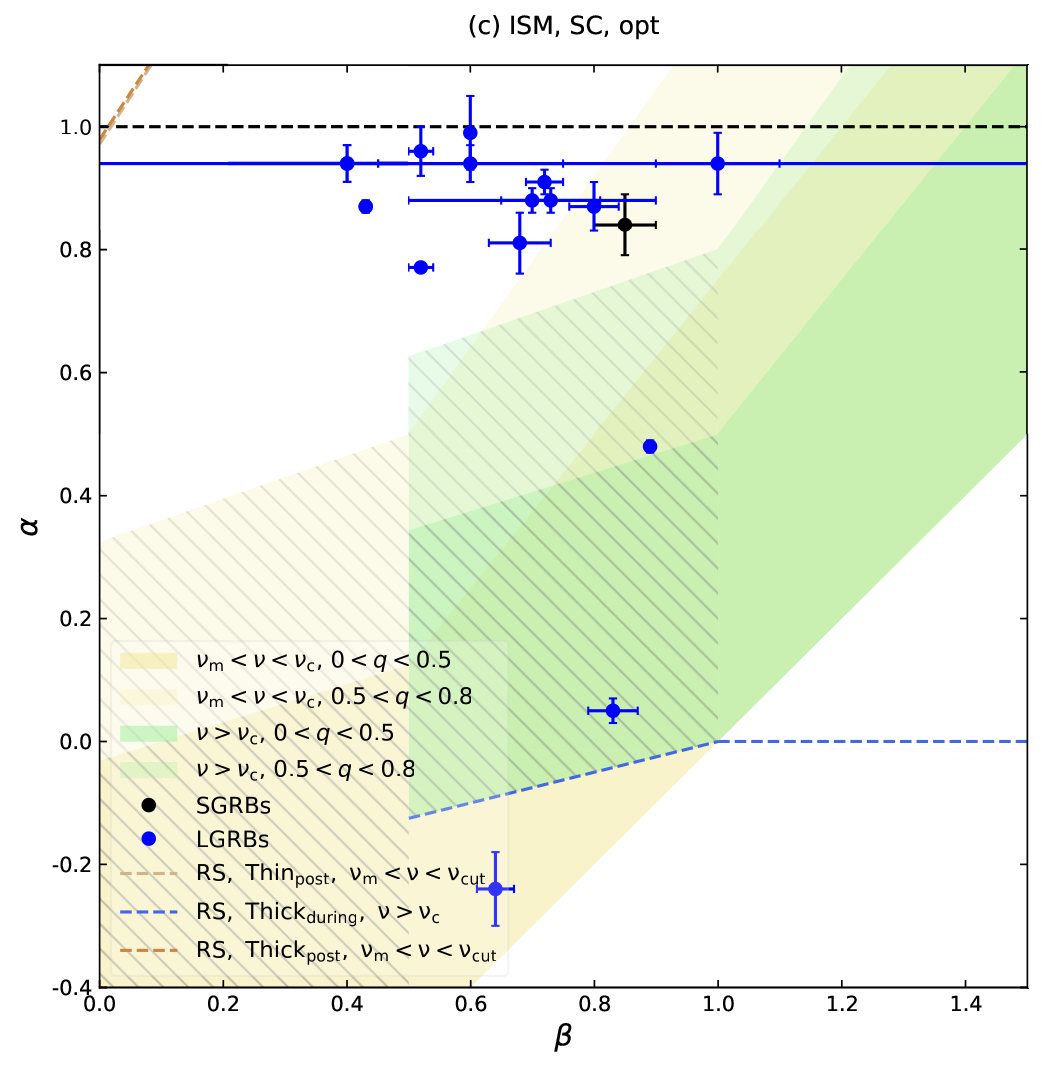}
\includegraphics[trim=0 0 10 10, angle=0,width=0.24\textwidth]{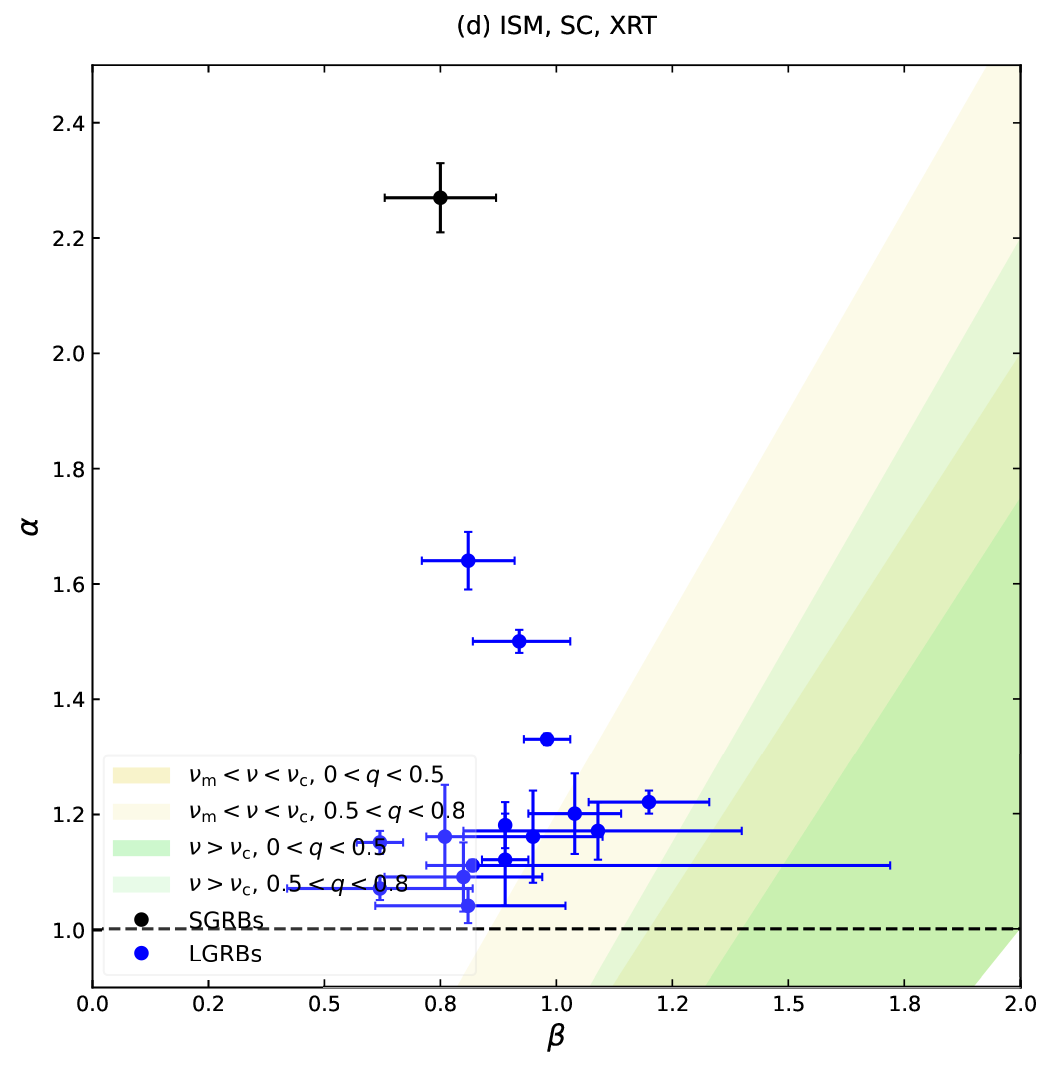}\\
\includegraphics[trim=0 0 10 10, angle=0,width=0.24\textwidth]{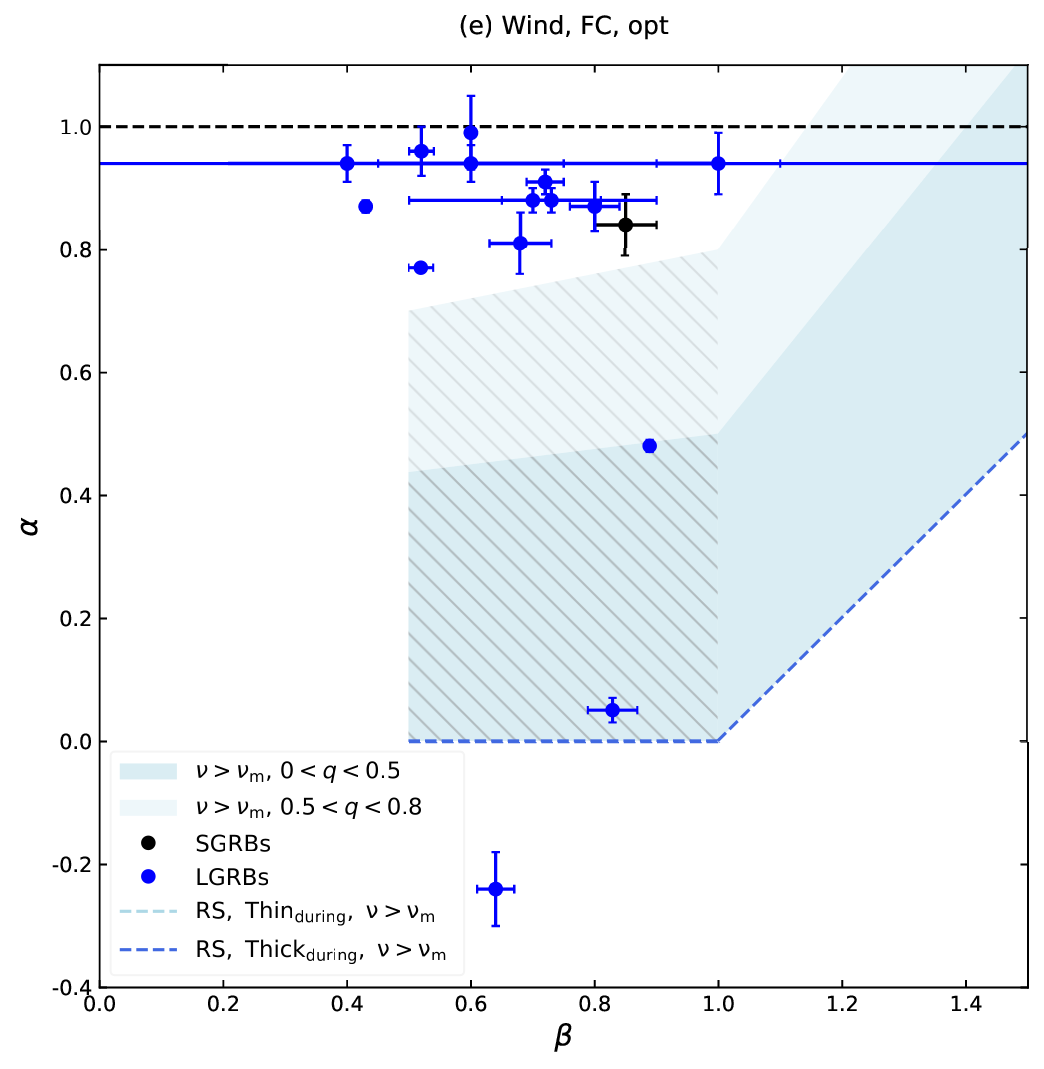}
\includegraphics[trim=0 0 10 10, angle=0,width=0.24\textwidth]{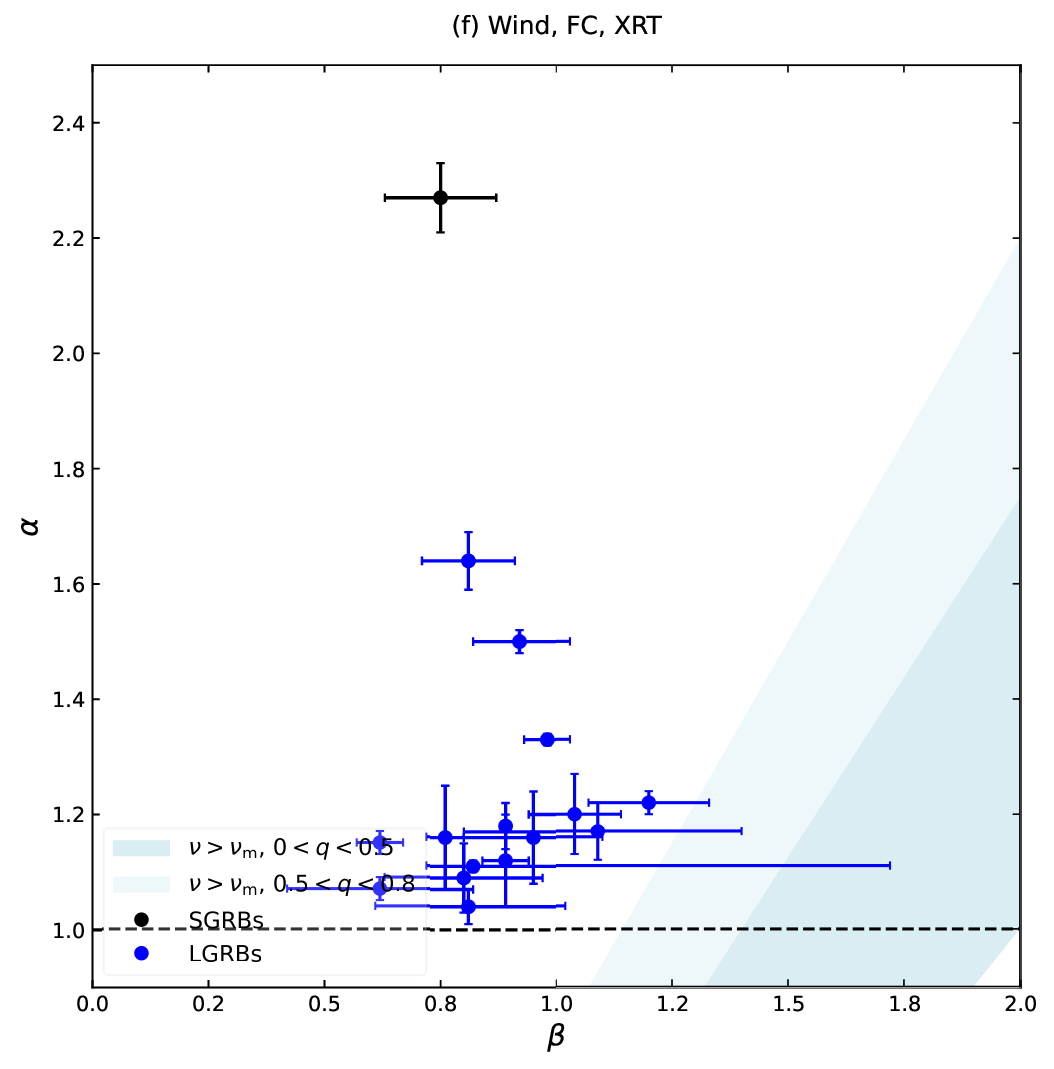}
\includegraphics[trim=0 0 10 10, angle=0,width=0.24\textwidth]{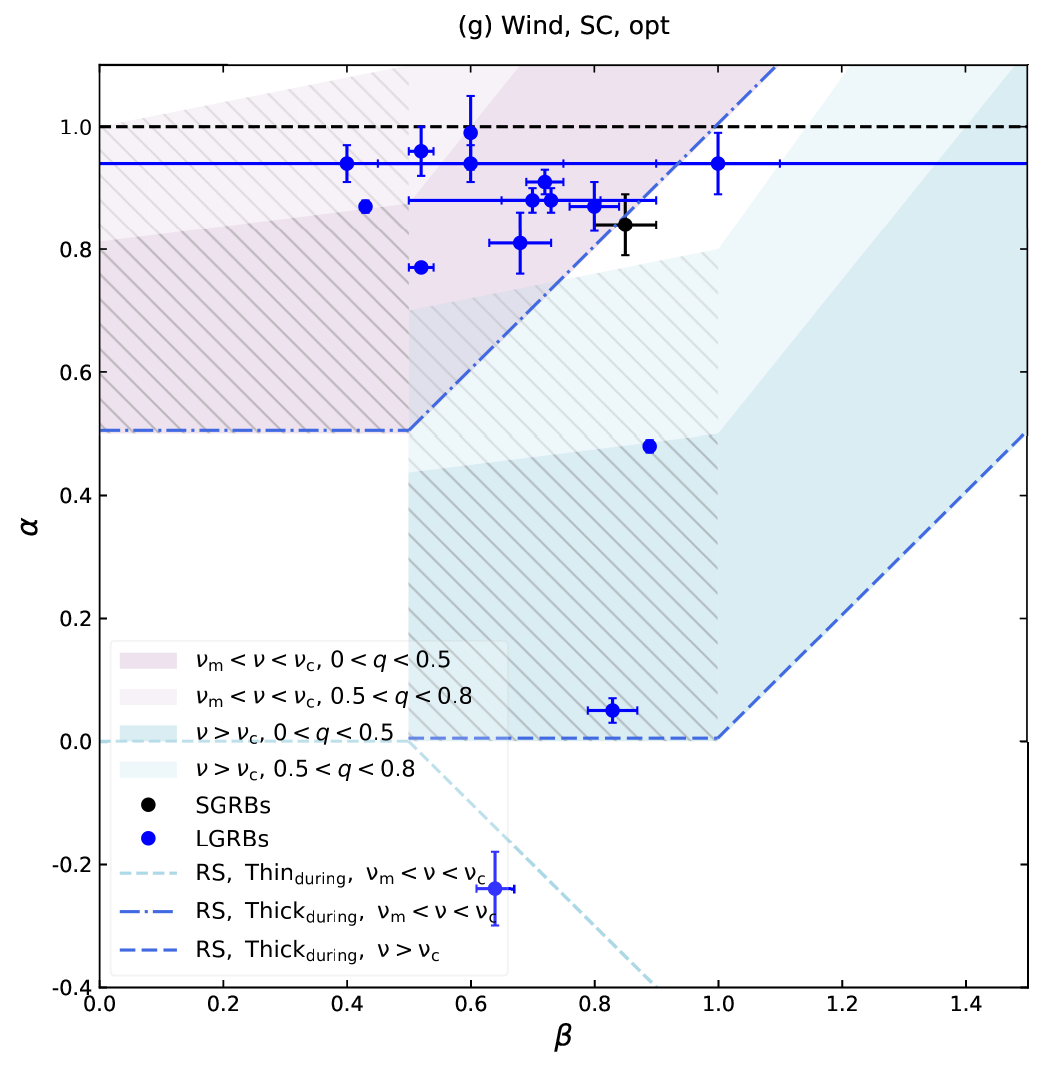}
\includegraphics[trim=0 0 10 10, angle=0,width=0.24\textwidth]{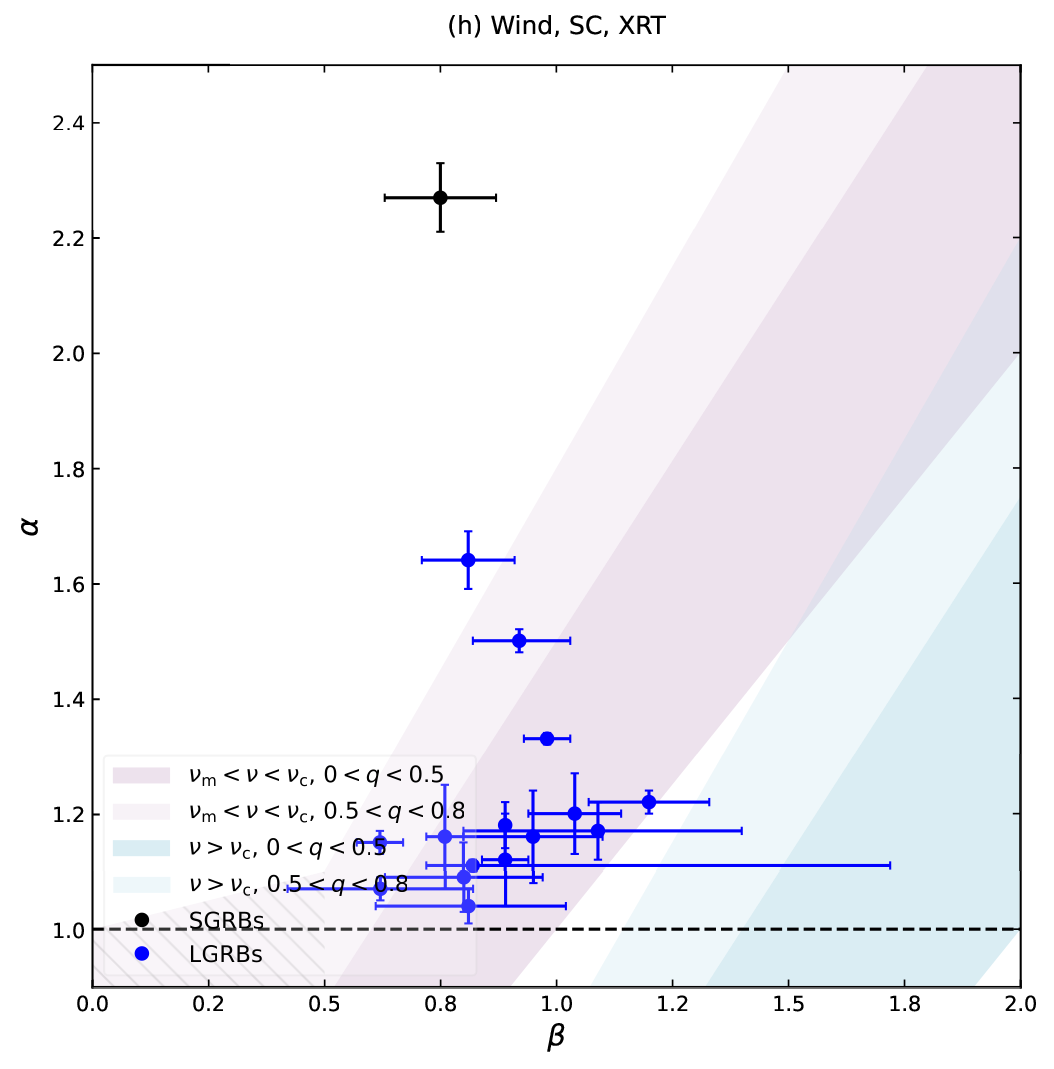}
\caption{Closure relations for dataset~3 with plateaus detected only in the optical band, in the same format as dataset~1.
\label{MyFig3}}
\end{figure*}

\begin{figure*}
\centering
\includegraphics[trim=0 0 10 10, angle=0,width=0.24\textwidth]{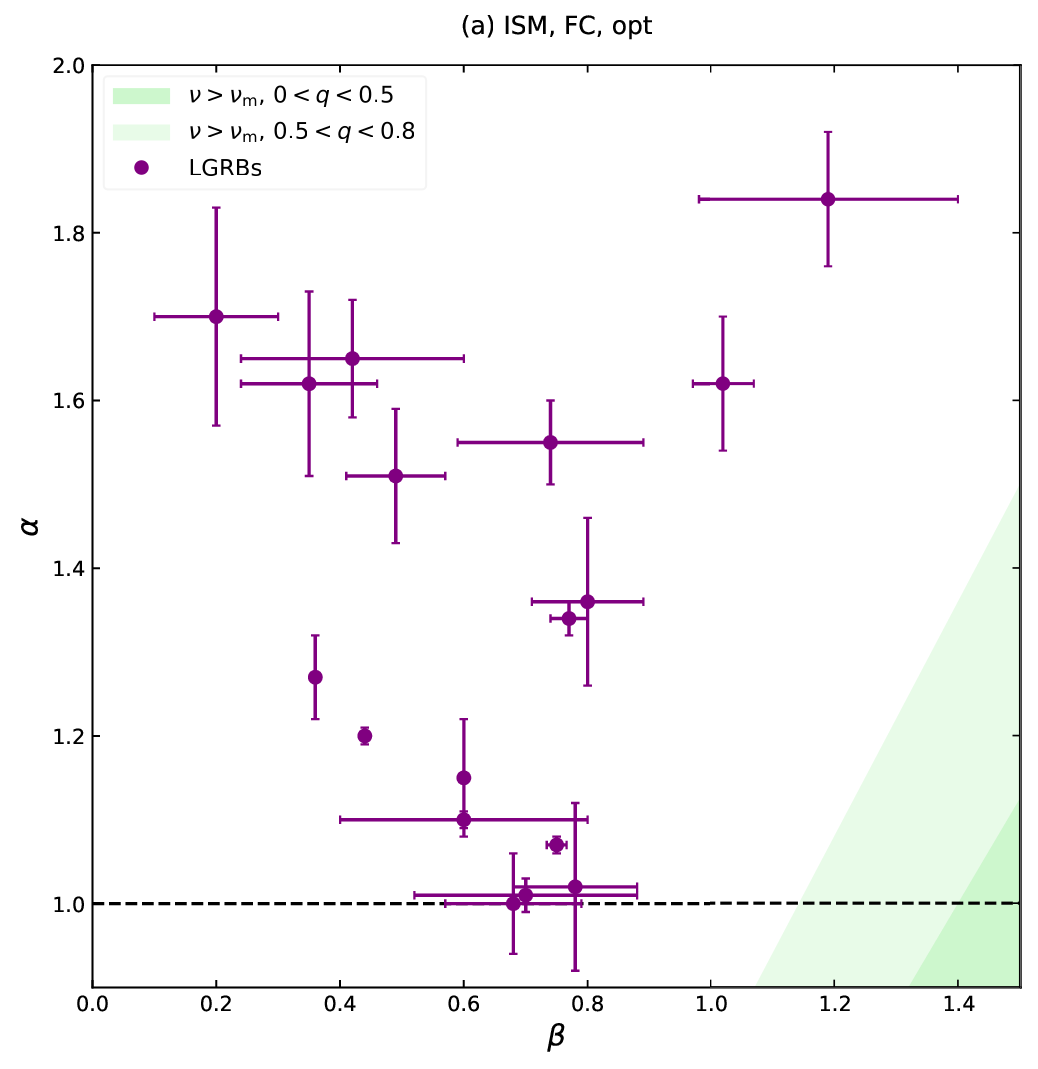}
\includegraphics[trim=0 0 10 10, angle=0,width=0.24\textwidth]{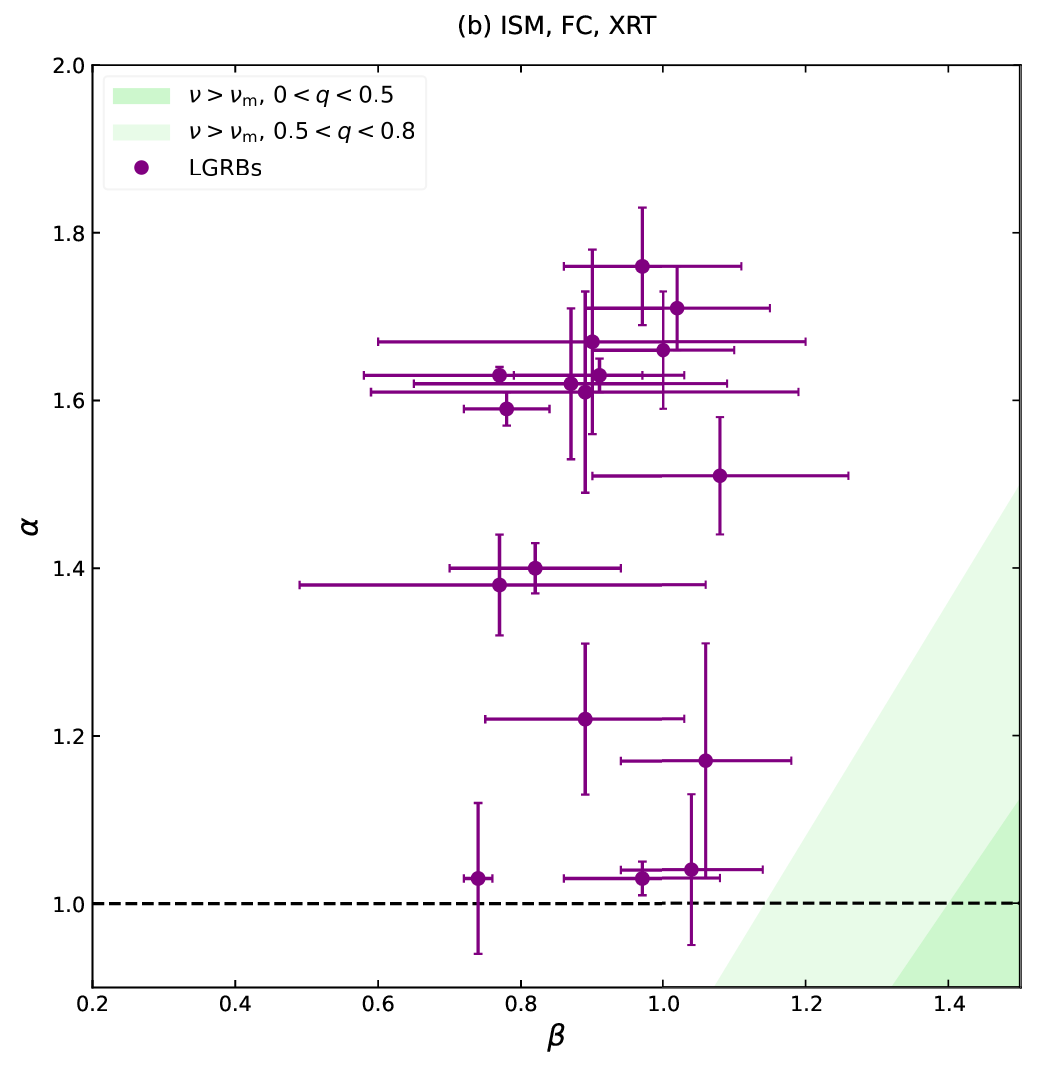}
\includegraphics[trim=0 0 10 10, angle=0,width=0.24\textwidth]{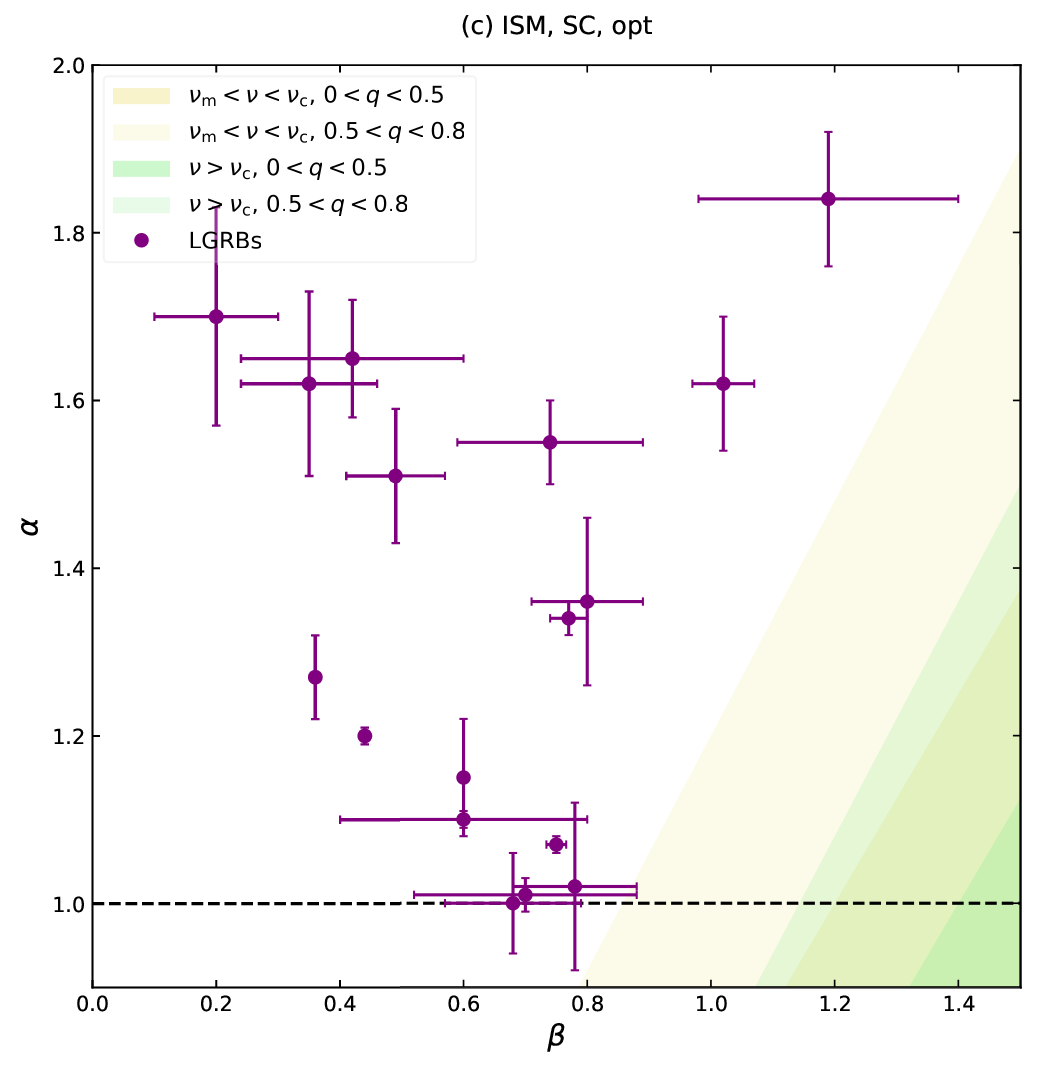}
\includegraphics[trim=0 0 10 10, angle=0,width=0.24\textwidth]{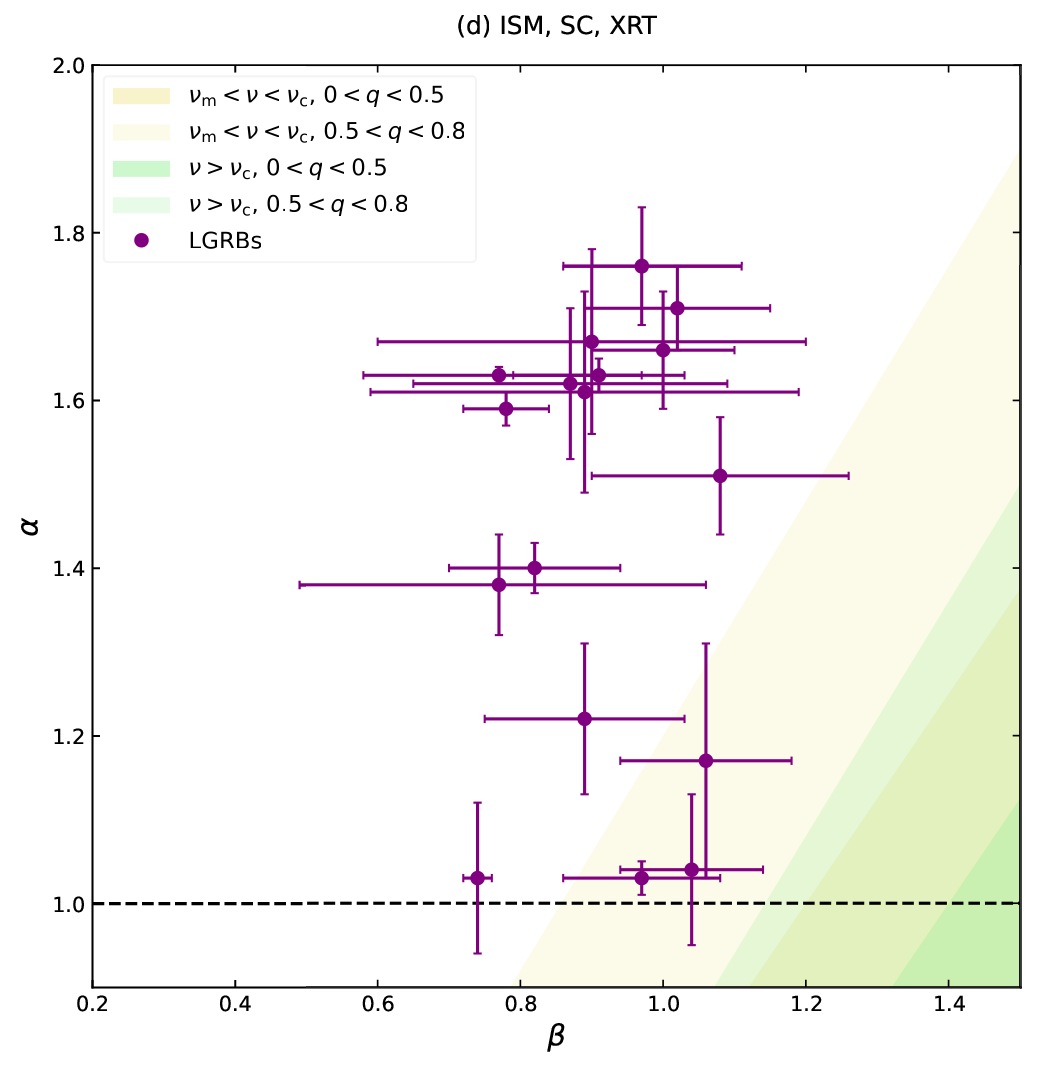}\\
\includegraphics[trim=0 0 10 10, angle=0,width=0.24\textwidth]{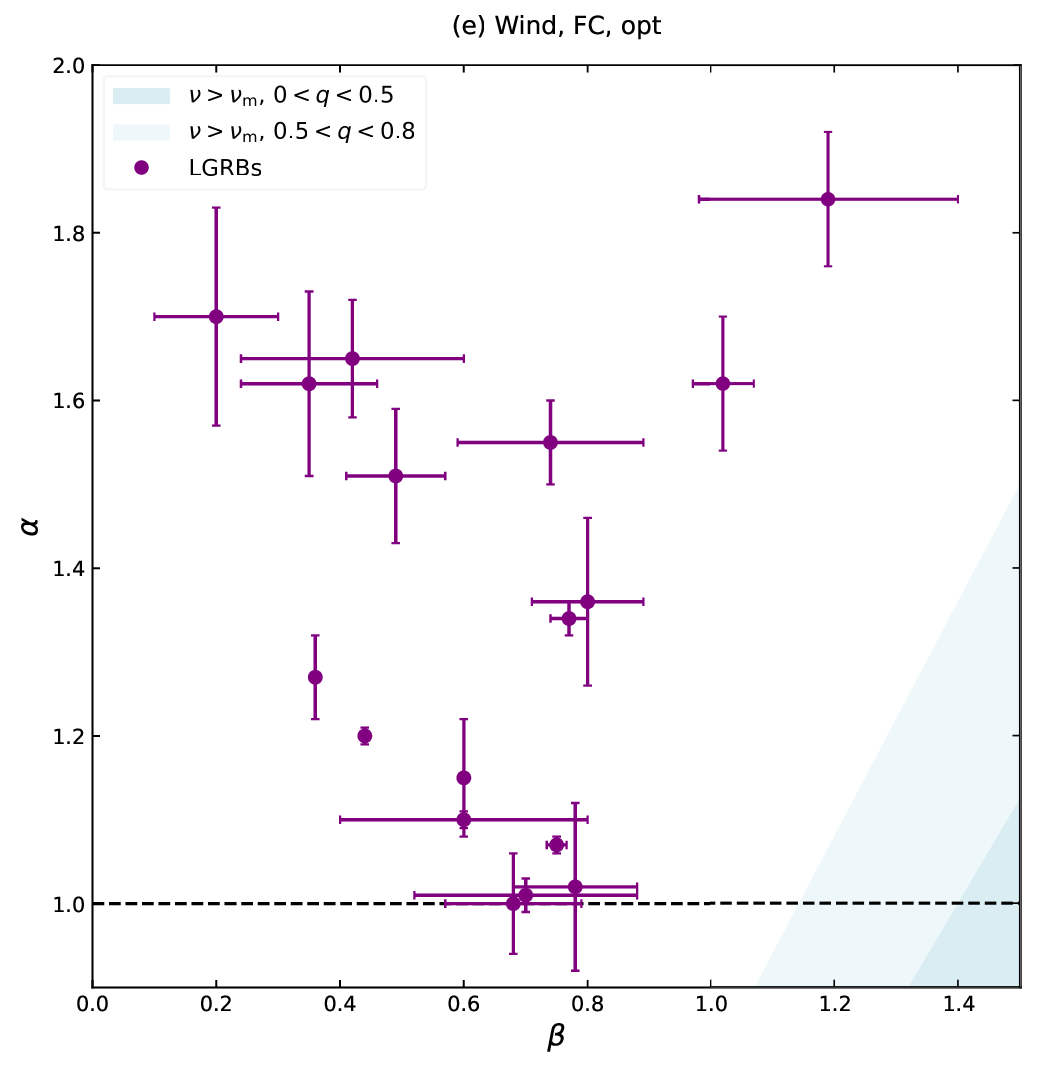}
\includegraphics[trim=0 0 10 10, angle=0,width=0.24\textwidth]{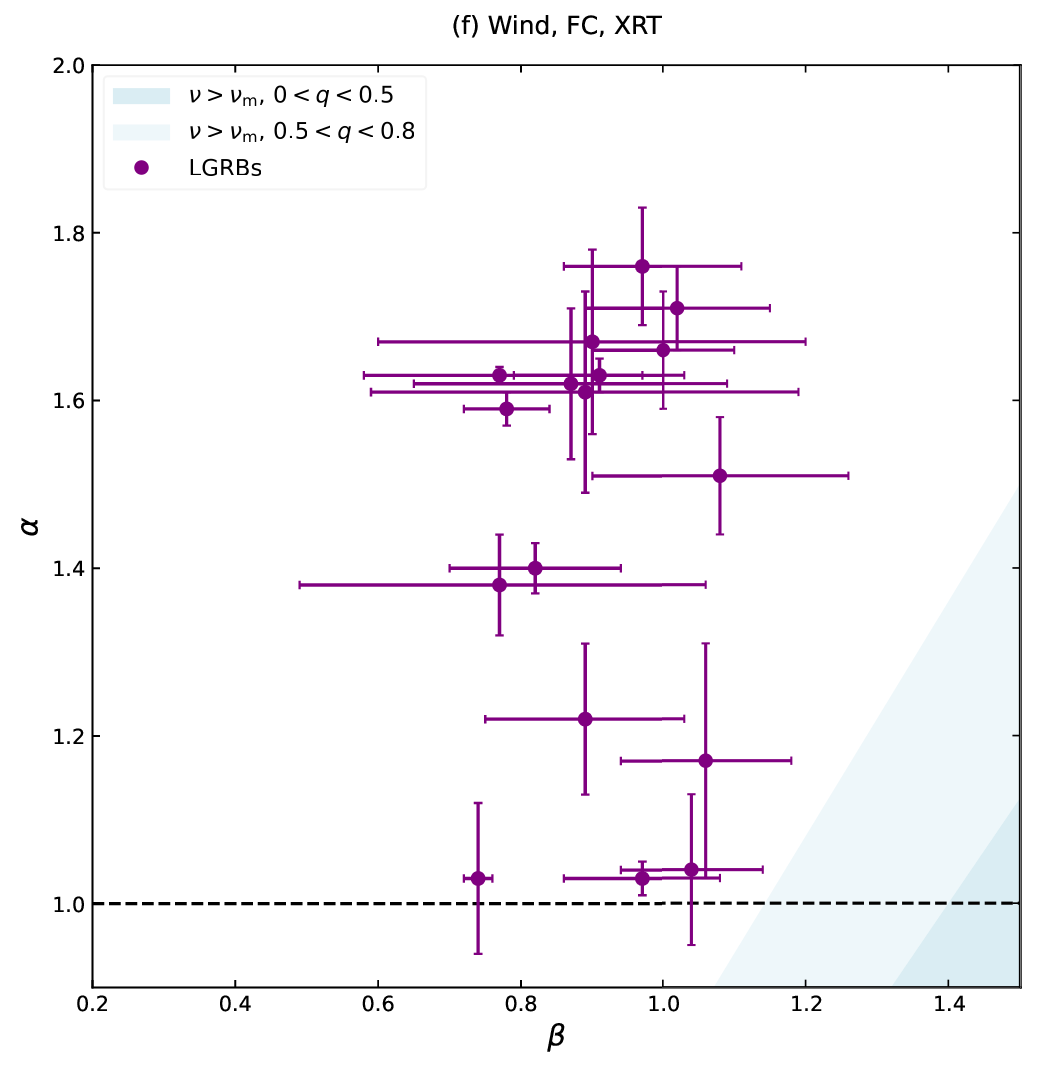}
\includegraphics[trim=0 0 10 10, angle=0,width=0.24\textwidth]{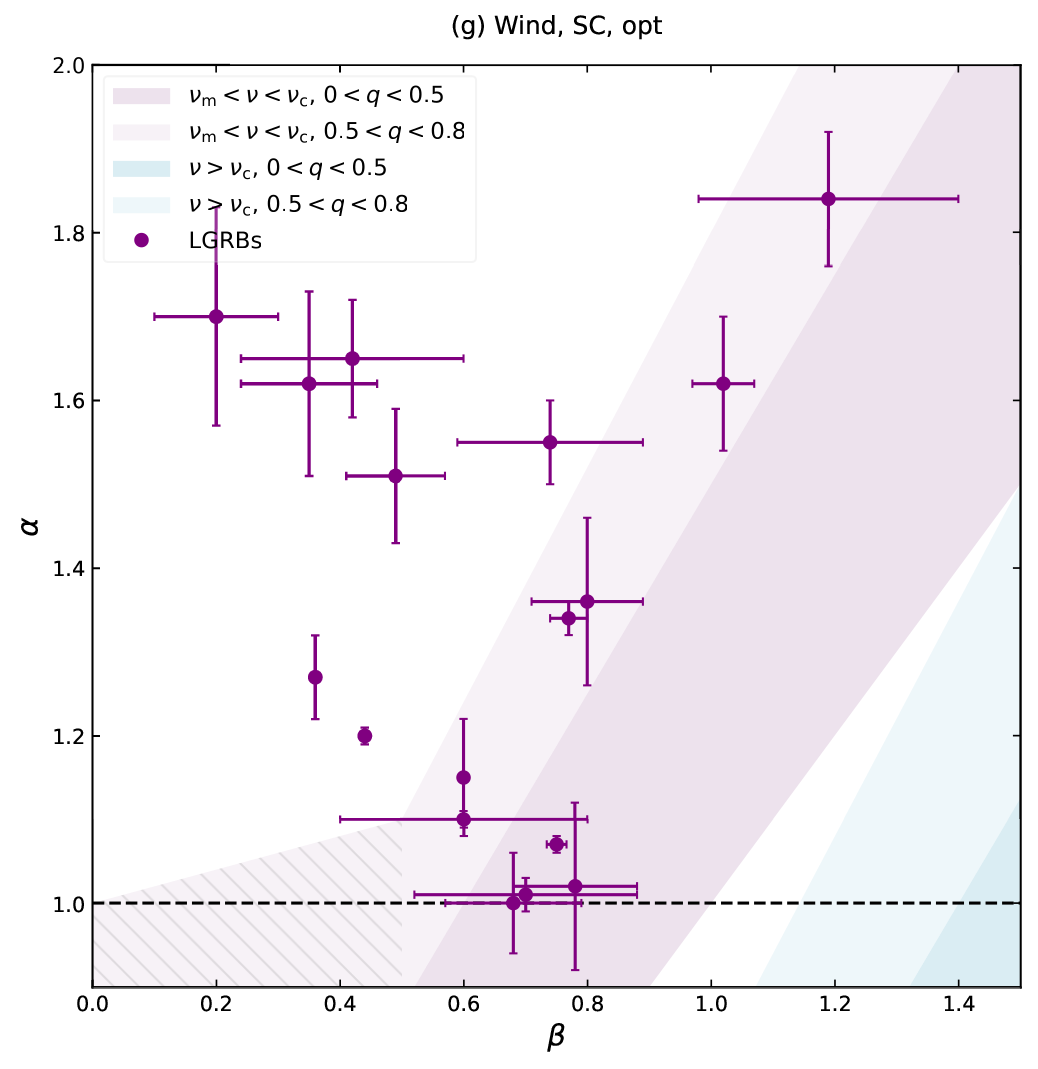}
\includegraphics[trim=0 0 10 10, angle=0,width=0.24\textwidth]{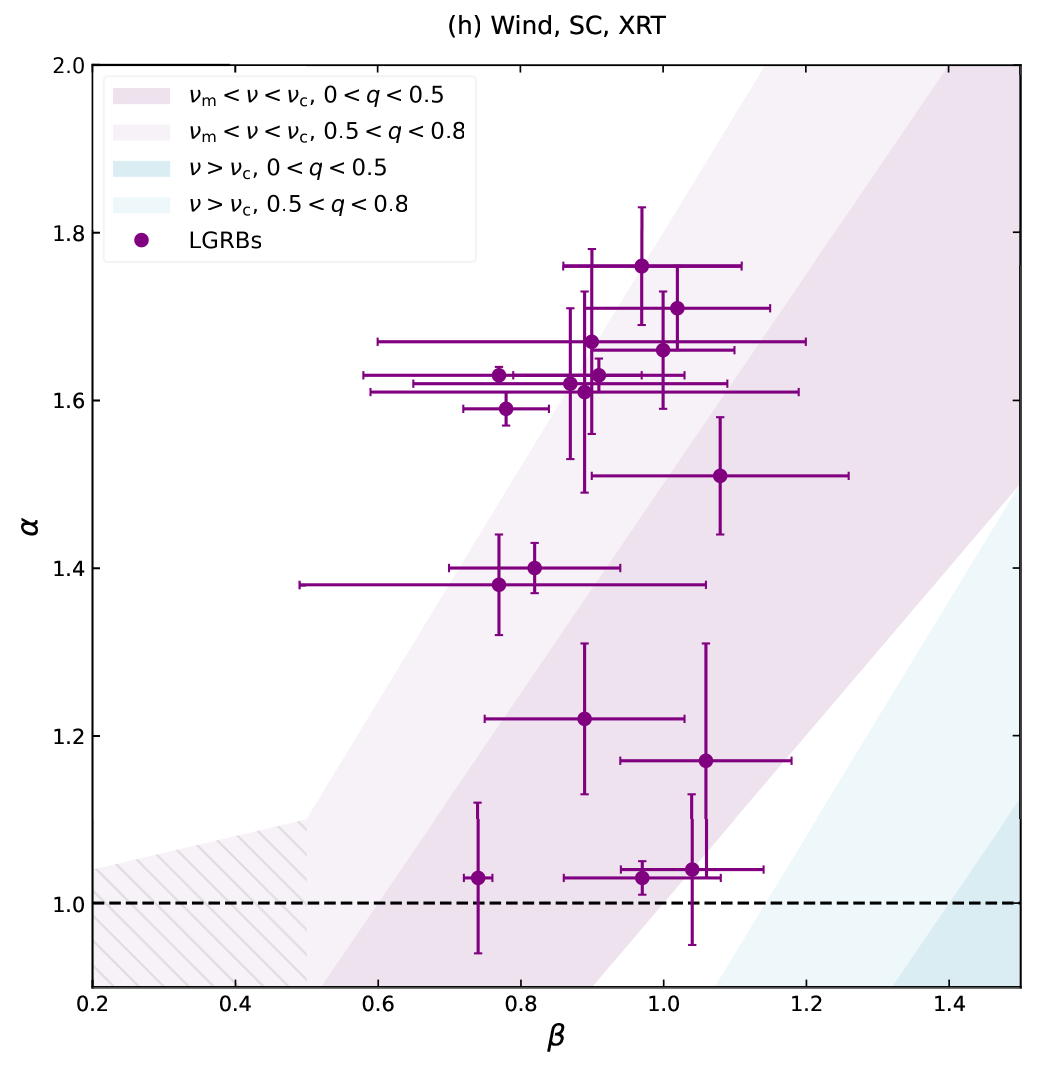}
\caption{Closure relations for dataset~4 without detected plateaus, in the same format as dataset~1.
\label{MyFig4}}
\end{figure*}

In Figures.~\ref{MyFig1}--\ref{MyFig4}, we present the temporal and spectral indices of the multi-band afterglows for the four datasets, together with the closure relations derived from Table~\ref{MyTabE}. The dashed black lines indicate \(\alpha = \pm1\), which are consistently adopted as the criterion for identifying plateau phases throughout this work. The regions corresponding to the injection parameter \(q\) within the ranges (0, 0.5) and (0.5, 0.8) are distinctly shaded in darker and lighter colors, respectively. The hatched areas within these regions represent the scenarios in which the electron spectral index satisfies \(1 < p < 2\); regions corresponding to \(p > 2\) remain unhatched. The green, yellow, blue, and purple shaded areas denote the results for the cases of the ISM fast-cooling (FC), ISM slow-cooling (SC), wind FC, and wind SC, respectively. The data points plotted in each figure illustrate the temporal and spectral indices of the multi-band afterglows in the corresponding datasets. Each figure consists of eight panels, corresponding to the following cases: (a) ISM, FC, optical band; (b) ISM, FC, X-ray band; (c) ISM, SC, optical band; (d) ISM, SC, X-ray band; (e) wind, FC, optical band; (f) wind, FC, X-ray band; (g) wind, SC, optical band; and (h) wind, SC, X-ray band. 

Figure~\ref{MyFig1} presents the results for dataset~1, in which plateaus are observed in both the X-ray and optical afterglow light curves. In this figure, red and black data points correspond to the cases of LGRBs and SGRBs, respectively. We first examine the ISM FC scenario presented in panels~(a) and~(b). As shown in these panels, a significant number of data points lie within the hatched regions corresponding to \(1 < p < 2\), implying an apparent consistency with the ISM fast-cooling scenario. However, such a situation appears challenging, while theoretically possible, as it is rarely observed that GRB afterglows possess electron spectral indices within \(1 < p < 2\). Moreover, during the plateau phase, the external shocks have generally decelerated, indicating that most afterglows should be in the SC rather than the FC regime. Consequently, we only consider the region of \(p > 2\) in the ISM-FC case, and only 6 GRBs simultaneously satisfy the closure relations in both the optical and X-ray bands for \(p > 2\) for both \(q \in (0, 0.5)\) and \(q \in (0, 0.8)\). It should be emphasized that only when both the X-ray and optical plateaus of a GRB simultaneously satisfy the closure relations can the plateaus be confidently attributed to the external-shock energy-injection model. In contrast, regarding the results for the ISM SC scenario shown in panels~(c) and~(d), we observe that irrespective of the GRB type (long or short) or observational band (X-ray or optical), the majority of data points fall within the shaded regions corresponding to \(p > 2\) and \(\nu_m < \nu < \nu_c\), which are consistent with typical GRB afterglow observations. After performing statistical analysis, we find that for \(q \in (0, 0.5)\), 28 bursts satisfy the closure relations of the ISM-SC regime, while the number increases to 50 when extending the range to \(q \in (0, 0.8)\). This strongly suggests that the plateaus in dataset~1 are favorably interpreted within the framework of external-shock energy injection.

The results for the wind FC cases shown in panels~(e) and~(f) exhibit a similar distribution to those of the ISM FC cases, which is expected because the closure relations of the wind FC scenario are identical to those of the ISM FC case under the condition \(p > 2\) (see Table~\ref{MyTabE}). In the wind SC cases, the optical-band data (panel~(g)) predominantly fall within the hatched region corresponding to \(1 < p < 2\) and \(\nu > \nu_c\), with only a few points located in the region of \(p > 2\) and \(\nu_m < \nu < \nu_c\). In contrast, for the X-ray band (panel~(h)), the number of data points within the \(1 < p < 2\) region noticeably decreases, while more data points appear within the \(p > 2\) and \(\nu_m < \nu < \nu_c\) regime. From the statistics, 26 bursts satisfy the closure relations of wind slow-cooling for \(p > 2\) and \(q \in (0, 0.5)\), and the number increases to 27 when \(q \in (0, 0.8)\). 
Combining the analyses of the ISM and wind environments, we find that 47 bursts of dataset~1 simultaneously obey the closure relations in both bands under the conditions of the electron spectral index $p>2$ and the injection parameter $q\in (0, 0.5)$, and 69 of the dataset for $p>1$ and $q\in (0, 0.8)$, providing a strong support for the energy-injection interpretation. 

On the other hand, those plateaus that cannot be satisfied by the closure relations under the energy-injection framework may, alternatively, originate from other physical scenarios, such as off-axis viewing effects \citep[e.g.,][]{2006ApJ...641L...5E,2007ApJ...655..973K}. Furthermore, it is also possible that some of these apparent plateaus are merely artifacts resulting from the relatively broad criterion adopted for their identification (i.e., $-1 < \alpha < 1$). It is noteworthy that a subset of bursts simultaneously satisfy the closure relations of both ISM SC and wind SC scenarios, implying that the closure relations alone may not always serve as a definitive diagnostic for distinguishing between ISM and wind environments. Nevertheless, this ambiguity does not significantly affect our main conclusions. 

In Figure~\ref{MyFig2}, we present results for dataset~2, where plateaus are identified exclusively in the X-ray band. The optical data in panel~(a) deviate significantly from the shaded regions, whereas the corresponding X-ray data in panel~(b) lie consistently within them. This phenomenon persists in panels~(c) and (d), as well as in panels~(e) and~(f). A deviation from this trend occurs in the last pair: while the X-ray data in panel~(h) satisfy the relations as expected, panel~(g) constitutes the sole exception where the optical data also lie within the shaded regions. Accordingly, only 3 bursts in the ISM and 7 bursts in the wind environment (for $p>2$ and $q\in(0,0.8)$) remain compatible with the closure relations. This result implies that the majority of X-ray plateaus in dataset~2 are not primarily produced by external-shock energy injection. Consequently, alternative central-engine processes are plausibly required—most notably spin-down luminosity from a magnetar—which can naturally and efficiently power X-ray plateaus \citep[e.g.,][]{2014ApJ...785...74L}. Otherwise, a higher emission flux from other components, such as optical flares, must have dominated and thereby obscured the underlying optical plateau emission.

In Figure~\ref{MyFig3}, we show the results for dataset~3, in which plateau features appear exclusively in the optical band. The blue and black points correspond to LGRBs and SGRBs, respectively. In the ISM environment, as shown in panels (a) and (b), the data points, corresponding to both the optical and X-ray bands, deviate significantly from the closure relations that account for energy injection under the FC regime. However, in the SC regime, displayed in panels (c) and (d), a small fraction of bursts exhibit consistency with the theoretical expectations in both bands. Statistical analysis shows that 4 GRBs in dataset~3 are consistent with the closure relations in the ISM SC case for $q \in (0,0.8)$ and $p>2$. In the stellar-wind environment, the behavior in the FS case is generally similar to that in the ISM; nevertheless, in the SC regime, a comparatively larger number of bursts satisfy the closure relations simultaneously in both optical and X-ray bands. Specifically, we find 13 bursts consistent with the closure relations in the wind case, lending partial support to the external-shock energy injection scenario. On the other hand, this also implies that a substantial fraction of optical plateaus in dataset~3 may not originate from continuous energy injection into the forward shock. Recent studies have proposed that reverse-shock emission could also account for the formation of optical plateaus \citep[e.g.,][]{2019pgrb.book.....Z}. Accordingly, in Figure~\ref{MyFig3}, we additionally plot the closure relations corresponding to the reverse-shock scenario in the optical band, denoted by dashed blue lines in panels (a), (c), (e), and (f). However,  $\alpha$ and $\beta$ values of the optical plateaus remain largely inconsistent with these relations. Furthermore, \citet{2024ApJ...972..158C} suggested that a two-component jet configuration can naturally produce chromatic plateau behavior, and the early optical plateau observed in GRB~191221B is well interpreted within this framework. In addition, \citet{2024ApJ...966L..25X} demonstrated that optical plateaus may also be shaped by the superposition of primary afterglow emission and pair-cascade radiation ($e^{\pm}$) generated through $\gamma\gamma$ interactions within the relativistic jets. Of course, if the optical plateaus indeed originate from external-shock energy injection, then an additional, brighter emission component, such as X-ray flares, must dominate the X-ray flux, thereby concealing the corresponding plateau signature in the X-ray band.

\begin{figure*}
\centering
\includegraphics[trim=0 10 10 10, angle=0,width=0.49\textwidth]{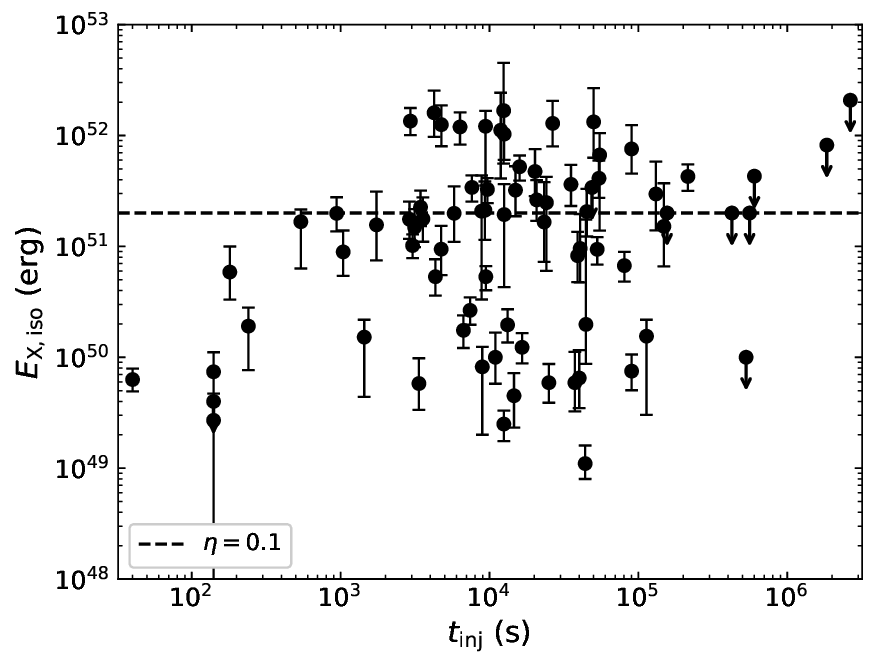}
\includegraphics[trim=0 10 10 10, angle=0,width=0.49\textwidth]{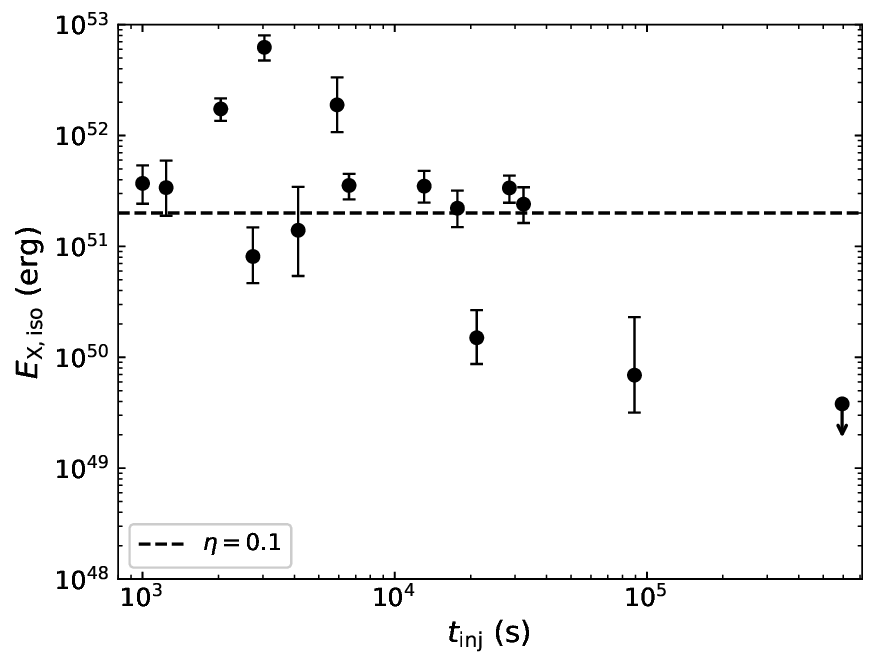}
\caption{The isotropic X-ray radiation $E_{\rm X,iso}$ for Dataset~1 (left pannel) and Dataset~2 (right pannel). The horizontal black dashed lines correspond to the $E_{\rm X, iso}=10\% E_{\rm rot}$.
\label{MyFig5}}
\end{figure*}

In Figure~\ref{MyFig4}, we present the results for dataset~4, which contains bursts without plateau features. In this figure, the purple points correspond to LGRBs. As can be clearly seen, the data points in both optical and X-ray bands deviate significantly from the theoretical closure relations, except for the wind SC case shown in panels (g) and (h). In these two panels, numerous data points fall within the shaded regions, and 11 bursts are found to be consistent with the closure relations. The reason why these bursts satisfy the relations in the wind SC case might be that their actual circumburst environments are not wind-like but closer to an ISM. Alternatively, the optical and X-ray afterglows of these events might originate from the dissipation of injected energy, rather than from the initial forward-shock radiation, and therefore do not manifest as observable plateau phases.

Overall, GRBs exhibiting multi-band plateaus statistically support the external-shock energy injection scenario, whereas this mechanism receives little support from those displaying single-band plateaus. For bursts exhibiting single-band plateaus, the underlying physical origins appear to be considerably more diverse and uncertain than those of multi-band plateaus. Since $\alpha$ measurements show slight variations among different studies, and spectral fitting is inherently affected by uncertainties related to the choice of temporal intervals, the energy-injection interpretation cannot be completely ruled out for GRBs whose $\alpha$ and $\beta$ of multi-band afterglow plateaus do not simultaneously satisfy the closure relations. More rigorous analyses, such as detailed numerical fits of the multi-wavelength afterglow emission, are therefore required to draw firmer conclusions. In addition, in this work, we adopted $\alpha$ values within the range of $-1$ to $1$ as the criterion for identifying plateau phases, which represents a relatively broad definition. A more restrictive threshold should yield slightly different statistical results. 

\subsection{GRB Central Engines}

In Figure~\ref{MyFig5}, we present the isotropic-equivalent X-ray energy, $E_{\rm X,iso}$, as a function of the injection timescale $t_{\rm inj}$ for dataset~1 (left panel) and dataset~2 (right panel), in order to roughly probe the nature of the central engine. The dashed line denotes the rotational energy budget available to a magnetar whose X-ray radiation efficiency is assumed to be $\eta = 10\%$, corresponding to an available energy of $2\times10^{51} \rm erg$. For dataset~1, 18 out of 75 bursts lie above this threshold, indicating that approximately $24\%$ of these events may be powered by a BH hyperaccretion rather than a magnetar. In dataset~2, 7 out of 15 bursts ($\sim47 \%$) exceed this limit. 

These estimates are derived under the assumption that both the magnetar rotational energy and the observed X-ray emission are isotropic. In the magnetar scenario, the total energy release is often considered to be quasi-spherical, particularly for SGRBs. For LGRBs, this assumption can also remain valid during the X-ray afterglow phase, especially at late times, such as during the plateau stage, because magnetar-driven outflows tend to become progressively more isotropic as the surrounding material is cleared away. Nevertheless, \citet{2014ApJ...785...74L} adopted a highly collimated configuration, and the realistic degree of asphericity could be between their assumption and that adopted in this work. Consequently, the derived results could vary depending on the degree of asphericity (see section~5.1 in \citet{2018ApJS..236...26L} for a detailed discussion).

\section{Conclusions and Discussion}

Plateau features are commonly observed in the afterglows of GRBs. In this work, we perform a statistical analysis of multi-band afterglow plateaus using a sample of 124 GRBs with both X-ray and optical observations. We examine the closure relations in the energy–injection scenario for the sampled GRBs. Our results show that most bursts exhibiting multi-band plateaus simultaneously satisfy the closure relations in both bands, thereby providing strong statistical support for the energy-injection model as a plausible explanation for GRB plateaus. However, we also find that bursts displaying single-band plateaus deviate from these relations, implying that additional physical mechanisms may be required to explain such cases. Furthermore, we estimate that approximately $24\%$ of GRBs with multi-band plateaus and about $47\%$ of those showing only X-ray plateaus might be powered by BH hyperaccretion, rather than by millisecond magnetar spin-down. 

Recently, \citet{2025JHEAp..4700384L} proposed a BH spin-down scenario powered by a magnetically arrested disk (MAD), which can naturally reproduce the observed anti-correlation and jointly model both X-ray and optical plateaus. Their analysis of a large multiwavelength sample suggests that BH spin-down provides a viable physical explanation for plateau emission.

GRB afterglows can also be detected in the radio band. \citet{2022ApJ...925...15L} analyzed 404 radio light curves and identified 18 events with plateau-like phases. Subsequently, \citet{2023MNRAS.519.4670L} tested these bursts against closure relations under different environments, cooling regimes, and both with and without energy injection. They found that the inclusion of energy injection does not necessarily improve the agreement with the standard fireball model. Within their sample, only GRB~071003 overlaps with our dataset; however, it does not exhibit a plateau in either the X-ray or optical bands. Therefore, no burst has yet been observed to display plateau phases simultaneously in the X-ray, optical, and radio bands. Coordinated multiwavelength campaigns are thus essential for probing the nature of multi-band plateaus.

Two plateaus have been observed in light curves of X-ray afterglows for some bursts, such as GRBs~070802, 090111, 120213A, 170714A, 160821B, 161217A, and 190114C. The nature of these bursts might be different. \citet{2018ApJ...854..104H} propose that they may originate from the merger of NSs. The NSs merge into a hyper-massive quark star (QS) and then collapse into a BH. In this scenario, the first plateau is powered by the latent heat released during the solidification of the hyper-massive QS. After the solidification, a high magnetic field is created and thus a magnetar-like central engine supplies significant energy to form the second plateau. More recently, \citet{2021ApJ...911...76X} proposed that a rapidly decreasing magnetic inclination angle of a nascent magnetar can lead to a two-stage energy injection process. This naturally results in a two-plateau structure in the X-ray afterglow light curve, which is well consistent with observations of GRBs~161217A and 190114A. On the other hand, the internal plateau can also form the two plateau phases. \citet{2018MNRAS.475..266Z} proposed that the first plateau, i.e., the so-called “internal plateau”, is produced by the spindown power from a supermassive magnetar. Then, the subsequent collapse of the magnetar into a BH results in a sharp drop after the first plateau. Fall-back accretion onto the newborn BH then powers long-lasting activity through the Blandford-Znajek process, producing the second plateau. This theory successfully explains the GRB~160821B. Similarly, \citet{2020ApJ...896...42Z} expanded the theory to LGRBs, and GRBs~070802, 090111, and 120213A have been explained well. Further theoretical investigations and multi-wavelength observations are required to clarify the true origin of these intriguing double plateaus. 

\section*{acknowledgments}
We thank Jiao-Zhen She and Yun-Feng Wei for the helpful discussion. This work was supported by the National Key R\&D Program of China (Grant No. 2023YFA1607902), the National Natural Science Foundation of China (Grant Nos. 12173031, 12494572, 12221003, and 12503052), and the Fund of National Key Laboratory of Plasma Physics (Grant No. 6142A04240201).

\end{document}